\documentstyle[12pt,aaspp4]{article}
\begin{document}
\tightenlines
%
%
\newcommand{\ea}{{et~al.}}
\newcommand{\IUE}{{\it IUE}}
\newcommand{\dmod}{$(m - M)_{0}$}
\newcommand{\logg}{$\log g$}
\newcommand{\logl}{$\log L$}
\newcommand{\lya}{\mbox{Ly$\alpha$}}
\newcommand{\lsun}{L$_{\sun}$}
\newcommand{\msun}{M$_{\sun}$}
\newcommand{\mv}{{M}$_{V}$ }
\newcommand{\rsun}{R$_{\sun}$}
\newcommand{\teff}{T$_{\rm eff}$}
\newcommand{\vsini}{$v \sin i$}
\newcommand{\rd}{Di\thinspace Stefano}
\newcommand{\re}{Einstein radius}
\newcommand{\er}{Einstein ring}
\newcommand{\ml}{microlensing}
\newcommand{\pl}{planet}
\newcommand{\mage}{magnification}
\newcommand{\lc}{light curve}
\newcommand{\lr}{lensing region}
\newcommand{\ev}{event}
\newcommand{\ec}{encounter}
\newcommand{\pop}{population}
\newcommand{\mo}{monitoring}
\newcommand{\bi}{binary}
\newcommand{\bis}{binaries}
\newcommand{\ob}{observation}
\newcommand{\ps}{planetary system}
\newcommand{\sy}{system}
\newcommand{\pr}{program}
\newcommand{\res}{resonant}
\newcommand{\ov}{overlap}
\newcommand{\cs}{central star}
\newcommand{\inm}{innermost planet}
\newcommand{\dbn}{distribution}
\newcommand{\shdn}{short-duration}
\newcommand{\evd}{evidence}
\newcommand{\zres}{zone for resonant lensing}
\newcommand{\sep}{separation}
\newcommand{\dtn}{detection}
\newcommand{\bl}{blending} 
\newcommand{\mt}{monitoring teams} 
\newcommand{\fut}{follow-up teams} 
\newcommand{\ft}{follow-up teams} 
\newcommand{\fss}{finite-source-size} 
\newcommand{\fsse}{finite-source-size effects} 
\newcommand{\cc}{caustic crossing}
\newcommand{\Asun}{A$_{\sun}$}
\newcommand{\Ajup}{A$_J$}
\newcommand{\Anep}{A$_N$}
\newcommand{\Nsun}{N$_{\sun}$}
\newcommand{\goesto}{\longrightarrow}
\def\mr{multiple repetitions}
\def\op{orbital plane} \def\lp{lens plane}
\def\os{orbital separation} 
\def\wo{wide orbit}
\def\w-o{wide-orbit}
\def\otn{orientation} \def\vy{velocity} \def\vt{$v_t$}
\def\smn{simulation}
\def\em{Earth-mass}
\def\jm{Jupiter-mass}
\def\rp{repeat}
\def\rpe{\rp ing \ev} 
\def\ss{Solar System}
\def\is{isolated}

\def\stacksymbols #1#2#3#4{\def\theguybelow{#2}
    \def\verticalposition{\lower#3pt}
    \def\spacingwithinsymbol{\baselineskip0pt\lineskip#4pt}
    \mathrel{\mathpalette\intermediary#1}}
\def\intermediary#1#2{\verticalposition\vbox{\spacingwithinsymbol
      \everycr={}\tabskip0pt
      \halign{$\mathsurround0pt#1\hfil##\hfil$\crcr#2\crcr
               \theguybelow\crcr}}}

\def\lapproxeq{\stacksymbols{<}{\sim}{2.5}{.2}}
\def\gapproxeq{\stacksymbols{>}{\sim}{3}{.5}}
\def\du{duration} \def\dca{distance of closest approach}
\def\stl{stellar-lens}
\vskip -.4 true in
\title{
A New Channel for the Detection of Planetary Systems Through Microlensing:
II. Repeating Events
}

\author{Rosanne \rd\altaffilmark{1},
Richard A. Scalzo\altaffilmark{2}}

\altaffiltext{1}{
Harvard-Smithsonian Center for Astrophysics,
60 Garden St., Cambridge, MA 02138; e-mail:  rdistefano@cfa.harvard.edu}

\altaffiltext{2}{Department of Physics, University of Chicago,
Chicago, IL 60637; e-mail:  rscalzo@rainbow.uchicago.edu}

\vspace{-0.15in}

\begin{abstract}

\vspace{-0.15in}
In the
companion paper we
began the task of systematically studying the detection of \pl s
in \wo s ($a > 1.5\, R_E$)
 via \ml\ surveys. 
In this paper we continue,
 focusing on
\rp ing \ev s.
Repeating \ev s are those in which a distant source is lensed
by two or more masses in the \ps . We find that, if all \ps s are similar
to our own Solar System, reasonable extensions of the present observing
strategies would allow us to detect $3-6$ repeating \ev s per
year along the direction to the Bulge.  Indeed, if \ps s with multiple
\pl s are common, then future monitoring programs which lead to the discovery  
of thousands of stellar-lens \ev s, will likely discover \ev s in which
several different planets within a single system serve as lenses,
with \lc s exhibiting multiple repetitions.
In this paper we discuss observing strategies to maximize the discovery
of all wide-orbit planet-lens \ev s.  We also  
compare the likely detection rates of \pl s in \wo s to those
of \pl s located in the \zres. We find that, depending on the
values of the planet masses and stellar radii of the lensed sources  
(which determine whether or not \fsse\ are important), and also
on the sensitivity of the photometry used by observers, 
the detection of \pl s in \wo s may be the primary route to the
discovery of \pl s via \ml . We also discuss how the combination of
resonant and \w-o\ \ev s can help us to learn about the distribution
of \ps\ properties (\S 6.1). In addition, by determining the fraction
of short-duration events due to planets, we indirectly derive
information about the fraction of all short-duration events that
may be due to low-mass MACHOs (\S 6.2).

\end{abstract}
\vskip -.2 true in

\section{Setting the Stage} 

Even before \ml\ \ev s were detected, it was realized that
\ml\ should provide a route to the discovery of distant \pl s.  
The first such route to be considered was through the discovery of
``resonant" events (Mao \& Paczy\'nski 1992, Gould \& Loeb 1992). 
These are \ev s in which the \cs\ of a \ps\ serves
as a lens, but the smooth standard form of the point-lens \lc\
is interrupted, for a time lasting typically from a few hours to a few
days, by a distinctive spike-like perturbation due to the presence
of a \pl . Such perturbations are expected for a fraction of 
stellar-lens \ev s when  the \pl\ orbits in an annulus between
$\sim 0.8-1.5\, R_E,$ where $R_E$ is the \re\ of the \cs .
  The observing teams have designed strategies to discover such
resonant 
\pl-lens \ev s. A second important channel has not been systematically 
considered
until recently: this is the channel provided by \pl s in 
wider orbits (\rd\ \& Scalzo 1997, 1998, \rd\ 1998a,b).
Such \pl s can also participate in \ml\ \ev s, either as    
single lenses, or as one of several lenses within the \ps\ that give
rise to a detectable ``repeating" \ev . 

The companion  paper was devoted to considering isolated 
\pl-lens \ev s in which the \lc\ exhibits a single continuous perturbation
from the baseline flux. This paper
focuses on \rpe s.  
By ``\rpe " we mean an \ev\ in which the track of a distant
star is lensed in a detectable way by more than one member of the
\ps .    
Together, the two papers provide a foundation
for observing programs to
 detect and study \pl-lens \ev s due to \pl s
in wide orbits. They establish that
\pl s in \wo s may provide an important discovery channel.
In this paper we also begin the task of   
considering likely relative rates of resonant \ev s and those due to 
\pl s in \wo s.  

In \S 2 we briefly discuss what we can learn from repeating events,
and also provide basic information about our notation and normalization
of event rates. 
In \S 3 we derive formulae for the general case of \ev s in which the track of 
a distant star crosses the \lr\ of more than one member of a \ps . In \S 4 we 
consider some specific \ps s, including our own \ss , to quantify how readily they 
would be detected as microlenses were they located in the  Galactic Bulge. 
\S 5 is devoted to a discussion of 
detection strategies that would optimize our ability
to detect \pl s in \wo s,  with special emphasis on \rpe s.    
In \S 6
we address the likely near- and long-term
results of implementing those strategies.
Finally, in \S 7 we summarize our results on the detection of \pl s in \wo s.

\section{Repeating Events}

{The idea behind the
definition of ``wide orbits" is  that, as the orbital separation between
a planet and the \cs\ increases, \cc\ \ev s become rare, and \ev s in which
the \pl\ acts as a more-or-less isolated lens become more common.
There is no one critical separation, beyond which all orbits can be
considered ``wide".} (See \rd\ \& Scalzo 1998) It is nevertheless useful
to have a specific criterion to use in calculations. We have
chosen $a_w,$ the inner boundary of the wide-orbit region, to be equal to $1.5\, R_E$. Our results on wide-orbit
\pl s are robust with respect to the choice of $a_w.$

When the 
\lc\ exhibits
detectable perturbations due to the presence of
more than one mass in the \ps ,
we refer to the \ev\ as a ``\rpe ".  
The most likely \rpe\ is one in which the \cs\ and the innermost
\w-o\ \pl\ serve as  lenses (\S 3). But more exotic events,
in which the path the source brings it near
to several planets, are also possible.

Because a key element of the definition of \rpe s is the detectability
of \lc\ perturbations, it is useful to define a \lr\
associated with each lens.  
Light from every star is, of course,
 deflected by every intervening mass. When we
use the term ``\lr," however, we refer to that portion of the lens plane
through which a source track must pass if 
lensing by a particular mass is to be {\it detectable}. 
For isolated (or, in the case of wide separations, almost-isolated)
lenses, the \lr\ is a disk surrounding the lens. It is convenient
to define the width, $w_i$, of the \lr\ to be the radius of
this disk.  
The value of $w_i$
depends on how \ev s are detected, as well as on the properties
of the lens and lensed source (see \S 2.2.1). The width of the \lr\ is important
because 
(1) the duration of the detectable perturbation from baseline
due to the $i\, $th lens is
proportional to $w_i;$ 
(2) the rate of detectable \ev s with $(k-1)$ repetitions
 scales as $\prod_{i=1}^k w_i,$.

\subsection{The Planet-Lens Event Menagerie}

Consider a star orbited by a \pl\ located a distance $a$ from it.
When this system serves as a lens, the likelihood of detecting
evidence of the \pl 's presence is determined by the value of $a.$
It is convenient to measure $a$ in units
of $R_E,$ the \re\ of the star. $w_{star}$ is the width
of the star's \lr\ (the region within which the track of the source
must pass in order for an event to be detectable); depending
on the sensitivity of the photometry
used to study the \ev , $w_{star}$ may be larger or smaller
than $R_E.$  

If $a$ is smaller than some minimum value, $a_c,$ the
large majority of lensing
\ev s will be indistinguishable from \ev s due to lensing by the
star alone. The value of $a_c$ depends on the mass ratio, $q,$ between
the \pl\ and star, and also depends to some extent on how good we
are at detecting deviations from the standard Paczyn\'ski \lc.
We will take $a_c$ to be equal to $0.8\ R_E,$ although in some cases it may
be significantly smaller.

When the planet is located within an annulus called the 
\zres ,  $a_c < a < a_w,$ with
$a_w\sim 1.5\, R_E,$ the \pl 's presence is most likely to be detected
through short-lived (${\cal O}(hours)$) perturbations of
\ev s in which the star serves as a lens.
(See Mao \& Paczy\'nski 1991, Gould \& Loeb 1992.)  
We will refer to such \ev s as 
``resonant" \ev s. Thus far, the search for \pl s has focused on
attempts to identify resonant \ev s. When  resonant \pl-lens \ev s are 
eventually discovered, 
it should be possible to determine
the values of $q$ and $a$ for many of them,
although there are degeneracies. (See, e.g.,
 Gaudi \& Gould 1997, and Gaudi 1997.)
The rate of detectable resonant \ev s falls sharply when the size of the source star,
as projected onto the lens plane, becomes large enough to be comparable to
the size of the caustic structure responsible for the perturbation (Bennett
\& Rhie 1996).

For $a > a_w,$ (i.e., for ``\wo s")
the \pl\ is most likely to act as an independent  
lens. Repeating events will dominate for $a < w_{star},$ 
where $w_{star}$ is the width of the star's \lr\ (the region within which the
track of the lensed source must pass in order for the \ev\ to
be detectable).
The rate of \rpe s falls (as $1/a$), but
\rpe s may play a significant 
role for separations out to several times
$R_E$. As is the case for resonant \ev s, the values of $q$
and $a$ can in principle be determined from study of the
\lc . Parameter extraction is generally more straightforward than it is
for resonant \ev s.    
 In particular, many \lc s can be well fit by separate point-lens
\lc s, and the others follow a simple
generalization (Di\thinspace Stefano \& Mao 1996).
Furthermore,  blending and \fsse\ affect each perturbation
in almost exactly the same way they influence point-lens \lc s.
This, combined with the fact that there are two lensing
\ec s, each with the same amount of blending, and of a source of
the same size, can make it easier to ferret out the physical
parameters that characterize the physical system.
As $a$ increases, 
isolated \ev s of \shdn\ 
dominate; from the \lc\ alone neither the mass ratio nor
orbital separation can be determined. 
\footnote{ We note that
for all \pl-lens \ev s, including isolated \ev s, 
 information about finite-source size and/or
blending can help to break degeneracies and also help us to learn
more about the lensing system than can be derived from the \lc\ alone
(see, e.g., \rd\ 1998a,b). For example,  
 finite-source-size effects can provide
further information (about the size of the caustic structure
for resonant \ev s and about the size of the \pl 's Einstein ring
for \w-o\ \ev s). Furthermore, if the \ps 's \cs\ is shining,
detailed study of blending, combined with high-resolution observations
carried out after the \ev, can provide further information about the
mass and state of evolution of the \cs.   
}
 
{\it In terms of the complexity of the \lc\
and the spatial regions giving rise to the \ev s, \rpe s provide
a bridge between resonant and isolated \ev s.}

\subsection{Notation}

Definitions specific to \rpe s are presented. In addition,
the notation
introduced and discussed in more detail in \S 2 of the companion
paper (\rd\ \& Scalzo 1998) is presented in abbreviated form, so that the
present
paper can be read independently.

\subsubsection{Photometric Sensitivity, Finite Source Size, and the Size
of the Lensing Region}

\def\aicp{the companion paper [\rd\ \& Scalzo 1998]}

As defined above,  
the \lr\ is that region in the lens plane through which the track of a
distant star must pass in order for there to be an observable
lensing \ev . The rate of \ev s involving an individual lens of mass $m_i$,
and the time duration of such \ev s,
is proportional to the geometrical width, $w_i$, of the \lr. 
The value of $w_i$ is related to the value of the \re\ of the $i$ th lens.
\begin{equation} 
R_{E,i}= \Big[ {{4\, G\, m_i\, D_S\, x\, (1-x)}\over{c^2}} \Big]^{{1}\over{2}}, 
\end{equation} 
where $m_i$ is the mass of the $i$ th lens, $D_S$ is the distance to
the lensed source, and $x$ is the ratio between the distance to the lens, $D_L,$
and $D_S$.  
The value
of $w_i$ depends on the photometric sensitivity.
If \fsse\ are negligible,
and if the minimum value of $A_{min}$, the peak \mage\ needed to 
reliably identify an \ev , is $1.34$ ($1.06$) then $w_i= 1\, R_E$
($2\, R_E$). When \fsse\ are negligible, $w_i = \sqrt{2} \,
\sqrt{[1 + {{1}\over{A_{min}^2 -1}}] -1 }\, R_E.$

Finite-source-size effects alter the width of the \lr\ in a manner
that depends on the size of the lens' \re\ relative to the size of the source as projected onto the lens plane.  Generally,
if the size of the source is slightly larger than the size of
the \er , the width of the lensing region can be increased.  
Because, however, the value of the peak \mage\ is decreased by \fsse ,
the \ev\ is not observable
when the source becomes too much larger than the \re .
For example, with $A_{min} = 1.06$  
and $R_S = 4\, R_E,$ $w_i\sim 4\, R_E$; if, however, 
$R_S =5\, R_E,$ then the \ev\ will not be detectable. 
More complete discussions can be found in
\rd\ \& Scalzo (1998), and \rd\ \& Scalzo (1997), 
\rd\ (1998b).
  
We define $n_i$ to be the ratio between the radius of the \lr\
of the $i$ th lens and its \re . That is, $n_i$ is the width of the \lr ,
expressed in units of the lens's Einstein radius:
$n_i = w_i/R_{E,i}$. Note that $n_i$ is not generally an integer.

\subsubsection{Encounters and Events}

When a \ps\ with several \pl s in \wo s serves as a lens, the track of the
source may pass through the region of influence of several lenses. 
When the 
source track passes within $w_i$ of lens $i$, we will say that an ``\ec"
is underway. We will use the word ``\ev" to refer to a source track 
(and the associated \lc) that experiences one or more \ec s.      

\subsubsection{Repeating, and Overlap Events}

When more than one \ec\ occurs, we dub the \ev\ a 
``repeating" \ev .
For many repeating \ev s, the orbital separation, $a,$ is larger
than the combined width of the \lr s, $w_1+w_2$. 
The \lc s associated with such \ev s exhibit genuine ``repeats";
i.e., between encounters, the \mage\ falls below the level at
which a perturbation from  baseline can be detected.  
As $a$ decreases below $w_1+w_2,$ the \lr s of the
two lenses begin to overlap.  Many \ev s which
involve \ec s with both lenses will give rise to \lc s in which the 
\mage\ does not fall below baseline between \ec s. The \lc s 
associated with these ``overlap" \ev s are similar
to others exhibiting repetitions, except that the perturbation due to
the first \ec\ is still noticeable when the second \ec\ begins.
We are almost guaranteed
to discover more than $1/2$ of all repeating overlap \ev s, even with no or
little change in detection strategy. 
As the value of $a$ decreases further, a larger fraction of the \ec s with
the smaller lens are part of \ev s in which the larger lens is
also encountered; for $a < w_1-w_2,$ this is true of all \ec s involving the
smaller lens. Most of the associated ``overlap" 
\lc s will exhibit a single connected
perturbation, with the structure in the wings largely
determined by the most massive lens, and the structure near one peak
largely determined by the less massive lens.
Although this class of ``overlap" \lc s may appear to be highly anomalous,
we should be able to fit them to the appropriate lensing model.
Since the duration of the single detectable perturbation from baseline
is governed by the size of the star's Einstein radius, these
events are long-lived (relative to \ec s with
an isolated \pl ),
we might be able to detect and identify $100\%$ of such overlap \ev s.
(See Figure 1.)  

\begin{figure}
\vspace{-2 true in}
\plotone{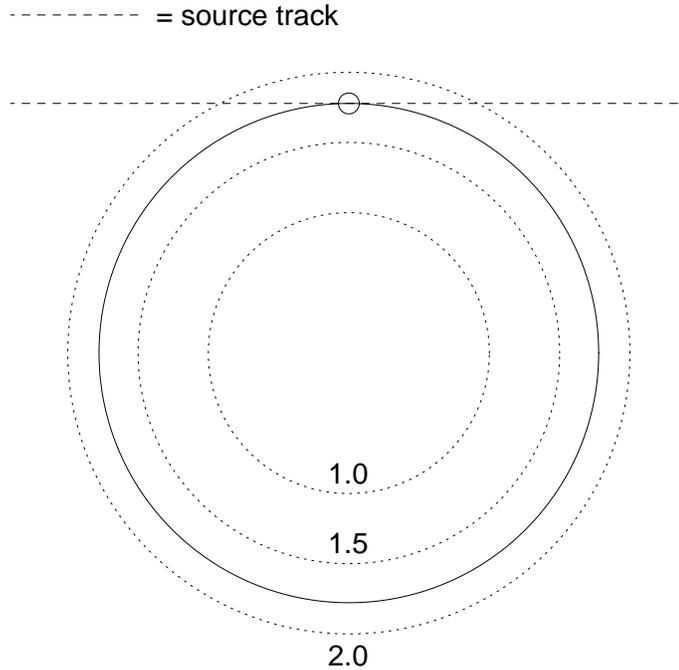}
\vspace{-2 true in}
\caption{The innermost \pl\ of the power-of-2 system used in the simulations
described in \S 4. 
The central star is taken to be of solar mass; we have assumed that the
lensed source and the lens are both in the Galactic Bulge ($D_S = 10$ kpc, 
$x = 0.9$). 
The dotted circles represent
the 1.0--, 1.5--, and 2.0\ $R_E$ rings for the central star.  The large solid
circle is the orbit (viewed face-on)
of the innermost wide planet, at a separation of 4.8 AU.  
The radius of the small solid
circle on the orbit is roughly equal to the Einstein radius of a 10-Jupiter-mass
object.
When $w_0 = R_E = w_{0,0},$  
there is a range of angles along which a source track can pass
through the Einstein ring of the planet without
also encountering the star. (See \S 2.2.5 for normalization information.)
When $w_0 =  2\, R_E = 2\, w_{0,0},$ 
any encounter with the
planet also results in an encounter with the star.    
}
\end{figure}

If $N$ is the number of \pl s, then 
in general we expect at most $N+1$ \ec s and $N$ repetitions.
In principle, if the source travels slowly and the \pl s orbit
rapidly, a single \pl\ could give rise to more than one \ec .
In practice this is extremely rare. Repetitions, however,
are not rare. As we will see, if most stars have \ps s similar to
our own, a data set containing the number of \ev s already observed 
along the direction to the Bulge should contain one or more \ev s
in which a repetition is due to lensing by a \wo\ \ps. As we will also
discuss, the detection of such \ev s can be optimized by frequent
sampling with sensitive photometry. 

\subsubsection{Normalization of Event Rates}  
All other things (such as the relative
velocities) being equal, 
the rate of \ev s is proportional to the width of the   
\lr . We therefore normalize the 
\ev\ rates by normalizing the width of the \lr s.
We define $w_{0,0}$ to be equal to $R_E,$ the Einstein radius of the
\cs . The width of every \lr\ is measured in units of $w_{0,0}$.
(Note that, if the photometry is good, $w_0,$ the width of the \lr\ of the
\cs\ may be larger that $w_{0,0}$; i.e., the normalized width of even the
\cs 's \lr\ can be 
different from unity.)     
At present, the rate of discovery of stellar-lens \ev s (using
a criterion which restricts the \lr\ to $0.75\, R_E$), 
is roughly $50$ per year. 
It is therefore
reasonable to assume $75-100$ \ev s of the type
we use for our normalization per year along the direction to the
Bulge.  
This means that, 
when we find  
that a particular detection
strategy leads to a rate of detectable 
\ev s of a certain type (e.g., \ev s with $2$ repetitions)
equal to $p\, \%,$ between $0.75\, p$ and $p$ such \ev s
could be discovered per year along the direction to the Bulge.
This  normalization is the same as that used
in the companion paper.

\subsubsection{Power-of-$k$ Planetary Systems}

A ``power-of-$k$" \ps\ is one in which 
the distance of the $i$ th \pl\ from the central star
is $k$ times the separation between the \cs\ and the $(i-1)$ th
\pl . For every \pl\ in the \zres\ in the power-of-$2$ ($3$) model,
we expect $\sim 10$ ($7$)
\pl s in wide orbits. \footnote{This assumes that the radial extent of a typical \ps\ is no larger
than $\sim 10^{17}$ cm.}

\subsubsection{Ongoing Observations}

Monitoring teams have found convincing evidence of over $200$ \ml\ \ev s.
(See, e.g., Alcock {\it et al.} 1997a,b (and numerous references therein); 
Udalski {\it et al.} 1997a,
1996, 1995, 1994, Ansari {\it et al.} 1996, Alard {\it et al.} 1995.)
These teams study tens of millions of stars every night,
looking for any deviations from baseline in those stars not known to be
intrinsically variable. Once a \ml\ \ev\ is thought to have started,
an alert is called, giving other observers around the world
an opportunity to carry out more frequent ($\sim$ hourly) monitoring. 
``Follow-up" teams have formed to take systematic advantage of the
\ml\ alerts.  
(See, e.g., Albrow {\it et al.} 1996, Alcock {\it et al.} 1997c,
Pratt {\it et al.} 1996a,b, Udalski {\it et al.} 1994b). The original  
monitoring and follow-up studies were of individual stars, each included 
on a template. More recently the benefits of using a differencing
approach which is sensitive to variations of stars not bright enough to
appear on the templates have been studied and are beginning to be implemented.
(See, e.g.,
 Ansari {\it et al.} 1997, Crotts 1996, Crotts \& Tomaney 1996, Tomaney  \&
Crotts 1996,   Han 1996, as well as the brief discussion in \S 5 of
the companion
paper [\rd\ \& Scalzo 1998].)    

\section{The Rate of Repeating Events}

\subsection{One Repetition}

Let $w_1$ ($w_2$) represent the width associated with lensing by the more 
(less) massive 
lens. 
We assume that the separation between the two lenses is wide--i.e., the
isomagnification contours associated with $A=1.34$ are distinct. As we
discussed in the previous section, this
means that if one lens is the central star and the other is
a \pl, $a > 1.5\, R_E.$ \footnote{We note that dynamical
stability of the \ps\ implies that two planets
will generally be separated from each other  
by significantly more than $1.5$ times the Einstein radius of the more massive \pl .}                      
The widths, $w_1$ and $w_2$, however can be smaller or larger than
the orbital separation, since their values are tied to issues of
photometric sensitivity. 

With $w_{0,0}$ defined as above, the detection rate
for repetitions involving lenses $1$ and $2$ is given by:
\begin{equation} 
{\cal R}_{1,2} = {{2}\over {\pi}} \, {{1}\over {2\, w_{0,0}}} 
\Bigg\{ \theta_{max} (w_1 + w_2) + \theta_{min} (w_2 - w_1) 
        + a\, \Bigg[ cos(\theta_{max})-cos(\theta_{min}) \Bigg] \Bigg\}
\end{equation}
In this expression, the values of $\theta_{max}$ and $\theta_{min}$ depend
on the value of the orbital separation, $a,$ as compared to the 
widths, $w_1$ and $w_2.$ 

\noindent For $a > w_1+ w_2 $,
\begin{equation} 
\theta_{max}=sin^{-1}({{w_1+w_2}\over{a}});
\, \theta_{min}=sin^{-1}({{w_1-w_2}\over{a}}).
\end{equation}

\noindent For $w_i+w_2 > a > w_1-w_2$,
\begin{equation} 
\theta_{max}={{\pi}\over{2}}; 
\, \theta_{min}=sin^{-1}({{w_1-w_2}\over{a}}).
\end{equation}

\noindent For $a < w_1-w_2 $,      
\begin{equation}  
\theta_{max}=\theta_{min}={{\pi}\over{2}}. 
\end{equation}

The limit $ w_1+ w_2 << a $  
corresponds to the case in which the separations
are extremely wide. In this case,
\begin{equation}
{\cal R}_{1,2} \cong {{2}\over{\pi}} {{w_1\, w_2}\over{a\, w_{0,0}}}
= n_1\, n_2\,
\Bigg[ {{2}\over{\pi}} {{R_{E,1}\, R_{E,2}}\over{a\, w_{0,0}}}\Bigg]. 
\end{equation}
The rate of repetitions is inversely proportional to the 
separation. Another key feature of this expression is the 
quadratic dependence of the width of the lensing region,
which is directly related to the photometric sensitivity.

The second extreme limit corresponds to the case when the separation
between the lenses is smaller than $w_1.$ In this case,
\begin{equation}
{\cal R}_{1,2} \cong {{w_2}\over{w_{0,0}}} = n_2\, 
\Bigg[{{R_{E,2}}\over{w_{0,0}}}\Bigg]. 
\end{equation}
That is, the rate of repeats involving lenses 1 and 2 is the same as
the rate of \ev s involving 2; all source tracks passing   
through the lensing region associated with 2 necessarily
pass through the lensing region associated with 1.

\subsection{Estimates} 

For the purposes of \ml , the most important difference between a
stellar binary and a \ps\ is that the \ps\ may contain several planets.
The presence of multiple \pl s provides more chances for \pl-lens
\ev s to occur. 
In addition, a small fraction of \pl-lens events (${\cal O}(0.001)$)
may exhibit clear evidence of the presence of multiple planets, as the
source 
travels close to the ecliptic of the \ps . 
A naive generalization of Eq.\ 6 leads to the following
expression for the average probability of events exhibiting a single repetition.
\begin{equation}
P_2 = \frac{2 \, n^2}{\pi T_{sys}} \int_0^{T_{sys}}
   \left[ \sum_{i=1}^{N}
          \frac{\sqrt{q_i}}{|\vec{a}_i(t)|} +
          \sum_{i=1}^{N-1} \sum_{j>i}
             \frac{\sqrt{q_i q_j}}{|\vec{a}_i(t) - \vec{a}_j(t)|} \right]
   \, dt,  \label{naive-approx}
\end{equation}
where $T_{sys}$ is the time taken for the configuration of the \ps\
to approximately repeat, and $\vec{a}_i(t)$ is the position vector
of the $i\,$th \pl , expressed in units of the stellar \re .
Note that we are considering the regime $a >> w_1 + w_2;$ thus,
in deriving this expression, we have implicitly assumed that none of
the \pl s is close enough to the \cs\ for an
overlap \ev\ to occur. The first term corresponds to the rate of repeats 
involving the \cs\ and one \pl . The second term 
corresponds to the rate of repeats 
involving two \pl s.  
 
To develop a feeling for the numbers, we will consider a simple
model, a
power-of-$k$ model, in which the plane of the
\ps\ is coincident with the lens plane and each planet has the same mass
ratio $q$ with the central star.
Using
\( \int_0^{T_{sys}} dt |\vec{a}_j(t)-\vec{a}_i(t)|^{-1}
\sim (a_j^2 + a_i^2)^{-1/2},\) allows us to simplify 
Eq.\ (8) in a way that does not overestimate 
the rate of wide-orbit lensing \ev s. 
\begin{equation}
P_2 = \frac{2\, n^2 \sqrt{q}}{\pi\, a_1}
   \left[ \sum_{i=0}^{N-1} \frac{1}{k^i} +
          \sum_{i=0}^{N-2} \sum_{j>i}
             \frac{\sqrt{q}}{k^i \sqrt{k^{2(j-i)} + 1}} \right].
\end{equation}
\noindent where $a_1$ is the distance between the star and the first
wide planet.
{For the power-of-2 model the value of the first term in
the above equation is approximately $2$, and that of the second term is
roughly $1.86\sqrt{q}.$
The rate of repeating \ev s
in which the star and one \pl\ each serve as a lens is 
$\Big[(4\, n^2 \sqrt{q})/{(\pi\, a_1)}\Big]$. If $q=0.001$ and $a_1=2,$ this
becomes $0.02\, n^2.$
\footnote{We note that, if $n$ approaches $2,$
then a \pl\  in an  orbit with $a=2$ would be detected through overlap \ev s;
$a$ would need to be larger for the approximation $a >> w_1+w_2$       
to apply. 
} 
The observed ratio between the rate of repeats involving 
both the star and one \pl , to the rate of stellar-lens \ec s is $0.02\, n.$
The ratio of such repeats to isolated \pl-lens \ec s is roughly 
$\Big[({2\, n})/({\pi\, a\, N})\Big],$ 
where $N$ is the number of \pl s in \wo s.
Note that the contribution of repeating \ev s saturates at moderate
values of $N$, while the rate of isolated \shdn\ \ev s is 
proportional to $N$. 
The ratio of repeating \ev s in which both \ec s are due to \pl\ lenses, 
to those in which one component is due to the central star, is suppressed 
by a factor ${\cal O}(\sqrt{q})$.

\subsection{Multiple Repetitions}

Multiple repetitions can occur when the stellar system contains $3$
or more masses. 
To derive analytic expressions, 
we first consider lensing by a static \ps\ whose \op\ is aligned with
the \lp . The effects associated with the \pl s' velocities and with
changing the \otn\ of the \op\ are then briefly considered.
The analytic expressions allow one to
make intuitive predictions. In \S 4 we turn to numerical simulations
to derive
detailed results that can be checked against the predictions.

\subsection{Multiple Repetitions in the Static, Face-On Approximation}

If a source track crosses through the orbit of a \pl ,
the probability that it will cross within $w_i$ of the
\pl\ itself is approximately $[(2\, w_i)/(\pi\, a_i)]$.
(We assume that the \pl 's motion during the time the
source crosses its orbit can be ignored, and that
curvature effects are also unimportant.)
Define $P_{\bf X}$ to be the probability that the source track will encounter
a specific subset, {\bf X} = $\{\beta_1, \beta_2, \ldots, \beta_k\}$,
of the system's masses, producing an \ev\ with $k-1$ repetitions.
Considering the regime $a>>w_i+ w_j$, the probability of encountering these
objects, and only these objects, is
\begin{equation}
P_{\bf X} \cong {{w_0}\over{w_{0,0}}} \prod_{i \in {\bf X}, i \neq 0}
   \left( \frac{2 w_i}{\pi a_i} \right).
   \label{eq:approx2}
\end{equation}
if the central star is one of the masses encountered.  If the central star is
not encountered, then
a similar expression is derived, but the contribution is suppressed by 
a factor on the order of $\sqrt{q}.$ 
We note that the curvature of the innermost planetary orbit in the set
{\bf X} may be significant and can increase the probability
that a track will encounter this planet.
The probability of an \ev\ with $k-1$ repetitions is obtained by
summing over all distinct sets, $X.$ 

The expression above is useful because it clarifies the functional form of the
probability of observing a particular combination of $k$ objects:
generally a rough
proportionality to the ratio between the
width of each \pl 's lensing region and its distance 
from the central star. This
makes it simple to estimate the effects of varying the parameters of the
source-lens system.

\tightenlines

\begin{deluxetable}{llllll}
\scriptsize
\tablecaption{Predictions: 
wide-orbit lensing events for known and model systems}
\tablehead{\colhead{System} \tablenotemark{(2)} &
   \colhead{$P_1-P_\odot$} \tablenotemark{(3)} &
   \colhead{$P_2^{sun}$} \tablenotemark{(4)} &
   \colhead{$P_2^{no sun}$} \tablenotemark{(5)} &
   \colhead{$P_3^{sun}$} \tablenotemark{(6)} &
   \colhead{$P_3^{no sun}$} \tablenotemark{(7)} }
\startdata
Known systems: \\
Gl 229       & 19.9  & 0.7  & 0.0     & 0.0     & 0.0 \\
PSR B1620-26 & 10.0  & 0.4  & 0.0     & 0.0     & 0.0 \\
\hline
Model systems: \tablenotemark{(8)} \\
Power-of-2 & 26.0  & 2.2   & 0.037 & 0.017
   & $8.2 \times 10^{-5}$ \\
Power-of-3 & 14.7  & 1.7   & 0.013 & $7.1 \times 10^{-3}$
   & $1.2 \times 10^{-5}$ \\
Power-of-4 & 12.6  & 1.5   & $7.1 \times 10^{-3}$ & $4.5 \times 10^{-3}$
   & $3.7 \times 10^{-6}$ \\
\enddata
\tablenotetext{(1)}{
The computations that produced the results in this table were based on the
analytic approximations discussed in \S 3.
Only encounters with
wide planets were considered; the results were averaged over all inclinations
of the system to the line of sight, and over all angles of approach for the
source track. The
detectability threshold is $A_{min} = 1.34$ (separation of $1.0 R_E$).
No minimum \ev\ duration was required for detectability.}
\tablenotetext{(2)}{
The only systems included are those that 
fall into the regime for wide lensing for $D_S =
10$ kpc and $x =0.9$).}
\tablenotetext{(3)}{Rate (per year) of isolated events in which a planet is
encountered. (See \S 2.2.5 for normalization information.)}
\tablenotetext{(4)}{Rate of events in which two encounters
occur, and in which the central star is one of the objects encountered.}
\tablenotetext{(5)}{Rate of events in which two encounters
occur, and in which neither object is the central star.}
\tablenotetext{(6)}{Rate of events in which three encounters
occur, and in which the central star is one of the objects encountered.}
\tablenotetext{(7)}{Rate of events in which three encounters
occur, and in which none of the three objects is the central star.}
\tablenotetext{(8)}{In each of these systems, the planet closest to the sun
is in an orbit of radius $a = 4.8$ AU.  The system is cut off at a 
maximum possible separation of $10^4$ AU.  Thus the power-of-2 system has 12
planets, the power-of-3 system has 7 planets, and the power-of-4 system has 6
planets.}
\end{deluxetable}

\subsection{Orbital Inclination}

\label{sec.inc-effects}

In general
the plane of the orbit will be inclined relative to the line of sight.
Let $\alpha$ represent the angle between the normal to the lens plane and the
normal to the orbital plane.
An approach analogous to the one sketched above for the face-on case
can be used.
The polar symmetry of the system is broken, however, since the
projection of an inclined circular orbit on the sky is an ellipse; the
probability of multiple objects acting as lenses depends on the direction from
which the source approaches the lensing system.
For some directions of approach,
the range of impact parameters leading to detectable events is now smaller,
but a larger fraction of events will involve encounters with more than
one object, and so will appear to repeat.
The effect can be intuited by multiplying each factor of
$a_i$ in the above equations by
the geometrical factor 
\begin{equation}
G \propto \left( \
   \frac{1}{\sqrt{\sin^2 \theta + \cos^2 \theta \cos^2 \alpha}}
   \right).
\end{equation}
where $\theta$ is the angle the track of the source makes
with the semimajor axis of
the elliptical projection of any planet's orbit.
Averaging over all possible angles of approach, there 
is a net increase in the
number of repeating events.

As $\alpha$ tends toward $\pi/2$ and we view the \ps\ edge-on, most directions
from which the source can approach can lead only to isolated \ev s. For a small
swath of tracks, however, the probability that the source will pass through the
Einstein radii of several 
\pl s approaches unity. If the \pl s are of equal mass and \ev s can be detected
when the distance of closest approach is $w_i = n\, R_{E,i},$ 
then the probability of an \ev\ in which the \cs\ and all of the \wo\ \pl s
serve as lenses is roughly $n\, R_{E,i}/D.$ 
where $R_{E,i}$ is the \re\ of a \pl, and
$D$ is the distance between the two most widely separated objects in the \ps .
The probability of \ec s with all but one or all but two of the masses in
the \ps\ are of roughly the same size, typically ${\cal O}(0.001).$ 
Thus, when the \ml\ teams have carefully followed the progress of
thousands of \ev s, they may have found several in which the source track traces the
global structure of a planetary system that gives rise to several detectable 
\ec s. 

It is interesting to note that, if the planetary system has a well-defined 
ecliptic plane, then events in which more than one wide planet is detected
will typically be those in which there is a fairly close approach to
the central star. Griest \& Safizadeh (1997)
have pointed out that, when there happens to be a \pl\ with 
projected position located in the zone for resonant lensing, then for
approaches in which the distance of closest approach is less than 
$\sim 0.1\, R_E$, there is a near certainty of
detecting the planet's presence. Gaudi, Naber, \& Sackett (1998)
have built on this result, to show that if the physical separation
between the central star and any planet is within the boundaries set
by the values of the zone for resonant lensing, there is still a
very high probability of planet detection; furthermore, more than
one planet can occupy this zone.  Thus, we may imagine that
close approaches to the central star that happen to follow the ecliptic
will discover several planets in wide orbits, and all planets
with true positions within $\sim 0.6-1.6\, R_E$ of the
\cs .

The orbital inclination also influences the wait times between \ec s. 
Averaging over all possible
directions of approach should cause a net decrease in the wait times between
encounters for repeating events, and should also produce a dispersion in the
distribution of wait times.

\subsection{Velocity Effects}

\label{sec.vel-effects}

The planets that can be detected via \ml\ are generally far enough from
the central star that their orbital speed, $v_{orb},$ is low. If
$a=\mu\, R_E$ then
\begin{equation}
v_{orb} = 7\, {{km}\over{s}}\, \left( \frac{2}{\mu} \right)^{\frac{1}{2}}
   \left[ \left(\frac{M}{M_\odot} \right)
   \left( \frac{10\, \mbox{kpc}}{D_s} \right)
   \left( \frac{1}{x(1-x)} \right) \right]^{\frac{1}{4}}.  \label{planetvel}
\end{equation}
 This shows that the planets that can be discovered
through microlensing, and especially wide planets, tend to be orbiting with
fairly low speed.  Hence, the transverse speed $v_t$ of the source (relative
to the central star of the lensing system) is likely to be large enough
compared to $v_{orb},$ that the static approximation used above can provide a
good guide to the \ev\ probabilities. Nevertheless, there will be a small
number of \ev s in which the relative magnitudes and \otn s of the transverse
source velocity and a \pl 's orbital velocity will influence the
encounter probability or the characteristics of an observed event.
This can occur
when \vt\ is drawn from the low-\vy\ end of the \vy\ distribution,
particulary if $x$ is close to unity (or zero) and the central
star is massive.
For example, if the transverse velocity of the source is
small enough that $R_{E,i}/v_t$ is a significant fraction of the orbital
period of planet $i$, the event probability increases.  This is because the
source spends a nontrivial amount of time within $R_{E,i}$ of the planet's
orbit;  if the planet does not lie in the source's path at the beginning of
this interval, it may move to cross the source's path during the interval,
resulting in a detection.
In addition, \ev\ \du s can be significantly decreased or increased,
depending on the angle between the directions of \vt\ and $v_{orb}.$

\section{Simulations of Lensing by a Planetary System}

The analytic formulae and discussions of the previous section allow
one to carry out back-of-the envelope calculations to determine 
the salient characteristics of the 
results we might expect observations to yield for single-planet
systems, and even for more complex systems.   
Given orbital separations, the stellar and planet masses, 
the orbital eccentricities, and the
orientation of the orbital plane with respect to the lens plane (or
averaging over possible orientations), 
simple calculations can determine whether
the rate of isolated and/or repeating \ev s is large enough to
expect that any specific \ps\ could be detected. It is worthwhile,
however, to carry out simulations to test the detectability
of some specific model \ps s. These simulations have allowed us to 
(1) include all of the relevant effects without making approximations, 
(2) study subtle features of
the distributions of event durations and wait-times due to the
presence of multiple \pl s, and (3) systematically study
the effects on the probability of detection 
of different detection strategies and a variety of \ps\ properties. 

To carry out these
simulations we must choose model \ps s, even though
very little is known about \ps s with multiple \pl s.    
The first system we have chosen is the one we know best, our own Solar System.
There are two features of the Solar System that are important
for the simulations, and which can be more-or-less 
studied separately even in the context of the simulation of this single system.
The first feature is that our Solar System contains a \pl , Jupiter, 
that is most likely to be detected through  overlap \ev s.
The influence of Jupiter can therefore, to  a good approximation, be
measured separately through the rate and characteristics of
overlap \ev s. 
The second feature is that there are several \pl s in wider orbits, with
a range of masses extending down from that of Jupiter.
Because of its proximity to the Sun, (and, to a lesser extent, because of 
the fact that its mass is larger than that of the other \pl s beyond its
orbit), Saturn provides the dominant contribution to
repeating \ev s in which the \mage\ falls below baseline between \ec s. 
There are also a significant number of \ev s in which Jupiter is
detected through an overlap \ec, and Saturn is detected through
a subsequent repeat. Jupiter and all of the outer \pl s contribute to
the rate of isolated events, but those beyond Saturn are almost
exclusively discovered through such \ev s. Because their masses
are smaller than those of Saturn and Jupiter, however, the short time duration
of these \ev s provides a challenge to their detection.
The Solar System model allows us to determine what we would see if
if all lenses were \ps s like our own--or if all \ps s have
(1) one \pl\ likely to be detected through overlap \ec s,  and (2) one
located roughly twice as far from the \cs . 

Model systems that are more extreme, in that they posit the existence of
a relatively large number of massive planets, are  
(a) a power-of-3 model with seven Jupiter-mass planets, and 
(b) a power-of-2 model
with twelve Jupiter-mass planets. The power-of-2 model was introduced
(Bennett \& Rhie 1996)  to ensure that each \ps , regardless of the orientation
of its orbital plane with respect to the lens plane, would be
likely to harbor a \pl\ in the \zres . Our simulations of these systems
share the salient features of the Solar System simulations, but
are designed to mimic the conditions explored by Bennett \& Rhie (1996) to
maximize the probability of detecting a \pl\ in the \zres. Thus, our simulations
of lensing by these systems defines the necessarily large role
that \pl s in \wo s will have if most \ps s do indeed have \pl s in the
\zres.

In any real galaxy it seems likely that \ps s will exhibit a variety 
of properties--\pl\ masses, orbital separations.
The observing teams will therefore detect ensembles of
\ev s which are superpositions of those we derive here and, very likely,
of even simpler \ps s. 

In each of the simulations we carried out,  one model \ps\ served as the lens.
The projections
onto the lens plane of a large number
 of randomly selected source tracks were followed.  
 In all cases we neglected ``inner" planets, i.e.,
those whose physical spatial separation from the central star
was less than $\sim 1.5\, R_E$, where $R_E$ is the
Einstein radius of the star.  All of the orbits have been
taken to be circular, with the motion governed by Newton's laws.  For each
model planetary-lens system, we have placed the source population in the
Galactic Bulge ($D_S$ = 10 kpc),
and the lens at $x\, D_S$ = 9 kpc.  
The central star in each of these systems was
chosen to be of solar mass.

At the time when the source started moving along the track with some
velocity, $\vec{v_t}$, we
started the planets in motion, each at a randomly chosen orbital phase.
As time progressed, we tracked the position of the source and of each
of the planets.
We wanted to determine when the source track passed close enough to any mass
in order for there to be a potentially observable ``encounter", roughly
how long each such encounter would last, and how many encounters there would
be as the source traveled along a specific track. Because the details
of the magnification as a function of time are not needed to derive this
information, we used only the value of the projected separation between the
source and each planetary-system mass to determine whether an
encounter was in progress. Specifically, we asked whether the projected
separation was smaller than some pre-selected value,
$w_i=n\, R_{E,i}.$ 
As before, $i$ labels the masses in the \ps, with $i=0$ corresponding to the
star, and $i$ ranging from $1$ to $N$ for the planets.
The values of $n$ to be used were selected at the beginning of each simulation.
[$n=1$ ($n=2$) corresponds to a magnification of $1.34$ ($1.06$).] 

$10^7$ tracks were used to sample the events expected when the solar
system served as a lens, and
$3\times 10^7$ tracks were used to sample the
power-of-$3$ and power-of-$2$ models.  
Because even the larger numbers of tracks used for the 
power-of-$3$ and power-of-$2$ models do not provide 
sampling equivalent to that used for the Solar System, we have
normalized the results so that the effective linear density of
source tracks was the same for each \ps .  We have also smoothed the
power-of-$2$ and power-of-$3$ distributions to remove effects  
due to poor sampling. 
We note, however, that the finer 
features of the statistical distributions, i.e., those which become
apparent only as the number of \ev s  increases, are best seen in the plots 
of the Solar System distributions. 
 
We have used our simulations to derive in detail what happens for
each planetary system, and also to test the effects of varying
some of the event parameters and detectability criteria.

\subsection{Event Parameters}

The characteristics of the \lc s
are determined by the characteristics of the event.
In particular, we must consider (i) the orientation angle, $\alpha,$
between the plane of the orbit and the lens plane, and (ii) the transverse
speed, $v_t,$ of the source with respect to the central star of the planetary-system
lens.
We have therefore carried out some simulations in which we have varied $\alpha$ and $v_t,$ in
order to test their influence on the results.
In the most realistic of our simulations, $\alpha$ was chosen uniformly over the
interval $0-2\, \pi$, and $v_t$ was chosen from a Gaussian distribution
centered at $150$ km/s, with width equal to $50$ km/s.

\subsection{Detectability Criteria}
\label{detectcrit}

There 
are two key elements of detectability.
First, does an encounter last long enough to be
detected?
Even in principle,  an encounter cannot be detected
unless it 
is caught in progress during at least one observation.  
Reliable detection generally requires the \ev\  to   
last long enough to span at least the time interval
between two or more consecutive observations.
In some of our simulations, we have assumed that a minimum \ev\ duration of
one day is needed in order to reliably detect the first encounter between a
source track and the \lr\ of a lens.  This criterion 
can be achieved by the \mo\ teams in some fields,
using their present observing
strategy. 
In others simulations,
 we have
dropped the requirement of a minimum duration for the first event.
This more relaxed condition is appropriate to the follow-up teams;
since they can achieve hourly \mo , even an \ev\ lasting $8$ hours
(such as one likely to be due to an Earth-mass \pl), 
can be
readily identified if the follow-up teams  attempt to discover new \ev s.

The second key element of detectability is provided by the value of the
peak magnification: 
what is the minimum
peak magnification, $A_{min},$ required in order to reliably determine that
an \ev\ occurred?
When the observing teams started, they tentatively chose $A_{min}=1.34,$
corresponding to a distance of closest approach equal to $R_E$.
It turned out, however, that some apparent events with magnification above
$1.34$ but less than $\sim 1.58$ were due to stellar variability; this has
led the MACHO team, for example, to use $A_{min}=1.58$. It is likely, however,
that this condition can be relaxed as the continued study of the
same fields over time will allow for better identification and tracking of stellar
variability and decrease its possible contamination of our count of
true \ml\ events. In principle, the value of $A_{min}$ is set by the
photometric precision of the monitoring system.
If the photometry is good to the $1-2\%$ level, 
then smaller values of $A_{min}$ are achievable. 
 $A_{min} = 1.06$ corresponds
to a distance of closest approach 
approximately equal to $2.0 R_E$, and  $A_{min} = 1.02$
corresponds
to a distance of closest approach 
approximately equal to $3.0 R_E$. (To achieve this latter
value would require better photometry than is typical of even the
present-day follow-up teams.) 
We have carried out two types of simulations. In the first, we have assumed
that \ev\ identification was being done by the \mo\ teams; in these, we
assumed that the distance of closest approach needed for the
identification of the first \ec\ is at least as
small as $R_E$. In the second, we have assumed
that \ev\ identification was being done by the follow-up teams; in these, we
assumed that the distance of closest approach needed to be at least as
small as $2\, R_E$. 
We note that finite source size effects can make \ev s
detectable when the distance of closest approach is even larger.
(See \rd\ \& Scalzo 1998.) We
return to this point in at the end of \S 4 and
again in \S 6; the results presented in this section
  were derived
under the assumption that the size of the lensed source could be
neglected.

We label
the three sets of detectability criteria used in our simulations:
``A'', ``B'', and ``C''.  

\noindent Criteria A:  The first \ec\ must exhibit magnification greater than  
	$A = 1.34$ (source-lens separation less than $1.0 R_E$) for at least
	1 day in order for the lensing \ev\ to be detected.  
        After the detection of a first \ec , 
	subsequent \ec s can be
	detected when $A > 1.06$ (source-lens separation less than $2.0 R_E$),
	and are not subject to minimum duration requirements. 

\noindent Criteria B:  All lensing \ec s are detected when the 
	\mage\ becomes larger than $1.06$ (i.e., the 
        source-lens separation becomes smaller than 
	$2.0 R_E$). The first encounter must have a duration of at least 1 day
         in order to be detected,
	but after one \ec\ has been detected,
        there is no minimum duration required for the \dtn\
        of subsequent \ec s.

\noindent Criteria C:  All lenses are detected when the
        \mage\ becomes larger than $1.06$ (i.e., the
        source-lens separation becomes smaller than
        $2.0 R_E$). 
	No minimum duration is required in order
        to detect any encounter.

Because we want to make contact with observations, we have defined \ec\ \du s
and wait times to reflect the actual required monitoring times
during and between \ev s. 
The duration of an encounter is always defined to be the interval of time
during which the source was within $2\, R_{E,i}$ of a lens. 
\footnote{Note that if a \mage\ of $1.34$ was required for \ev\ detection,
the duration can nevertheless be 
(and for us, is) defined as the time during which
the \mage\ was above $1.06,$ since this information can readily
be extracted from the
\lc . This is because the sensitivity of the photometry is always significantly
better than the fractional increase in \mage\
required in order that the
\ev\ as a whole be reliably associated with \ml .}     
Wait times are defined as the time intervals between encounters. 

We have used the normalization of \S 2.
$P_1$ is the rate (per year) of source tracks that cross through the \er\ of a
single object. 
All rates reflect the percentage of \ev s of the given type
relative
to the rate of isolated
stellar lens \ev s
(with $A_{peak}>1.34$) in our 
simulations. 
\footnote{Note that the exact normalization described in \S 2
requires dividing by the number of \ec s, rather than by the number of
isolated \ev s in which the \cs\ serves as a lens. Because these two different
ways of computing the rates lead to similar results (using the number of
\ev s yields rates that are $\sim 2-10\%$ higher), and because
the number of \ev s provides a more straightforward comparison,
we have chosen to divide by the number of isolated stellar-lens \ev s.} 

$(P_1-P_\odot)$ is the rate (per year) of source tracks passing through
just one object, excluding cases in which the \cs\ is \ec ed. $P_{1,\ov}$
is the rate (per year) of source tracks which 
in which the influence of \ec s with $2$ lenses
is clearly
visible in the \lc, but the \lc\ exhibits 
just one continuous perturbation 
(i.e., once the \mage\ falls below 
the detectability limit, it does not again
rise above it).
$P_i$ is the rate (per year) of source tracks crossing through
the \lr s of $i$ objects. $P_{i,\ov}$
is the rate (per year) of source tracks passing through $i+1$ 
\lr s, but in which
the presence of two of the lenses (the \cs\ and the innermost \pl\ in a \wo ) 
is
detected in an \ov\ \ec ; the \mage\ rises above and falls below
the detectability limit  
only $i$ times.

\subsection{Results}

The simulations allowed us to compute the rates of both isolated
and repeating \ev s.  The rates of all events are provided in Table 2.
Figures 3-6 show the
distributions of
properties for those \ev s with \lc s exhibiting
two separate deviations from the baseline flux.

The primary focus of this section is on repeating \ev s in which the
\lc\ displays more than one deviation from baseline. It is important
to note that \ov\ \ev s involving just 2 lenses produce only a single
deviation from baseline; graphs showing the
distribution of durations for such \ev s are 
therefore included in
the companion paper (\rd\ \& Scalzo 1998). Although, from the
phenomenological perspective, such \ev s are ``isolated", conceptually
they are ``repeating", in that the track of the source does pass through 
the \lr s of 2 lenses.  Their importance can be assessed in two ways.
Judging by the rates, they are an important because, depending on the
spatial and mass distribution of \pl s, their frequency
is generally
comparable to that of other \rpe s. Judging by their
detectability, they are, in the short term, 
almost certain to play a larger role than
other \rpe s.
This is because they should be observable primarily
as apparently perturbed isolated stellar-lens \ev s , with no ``wait time"
between \ec s. In the long-term, observations of all 
\rpe s will be important. The relative numbers of overlap \ev s
and \rpe s exhibiting multiple rises from the baseline flux,
 will
 help us to learn about typical 
spatial distributions of \pl s within \ps s.
In the very long term, the observation of a small number of \lc s with
multiple repetitions will provide exciting sketches of distant \ps s.

\subsection{The Effects of Systematic Variation of $v_t$} 

\begin{figure}
\vspace{-1 true in}
\plotone{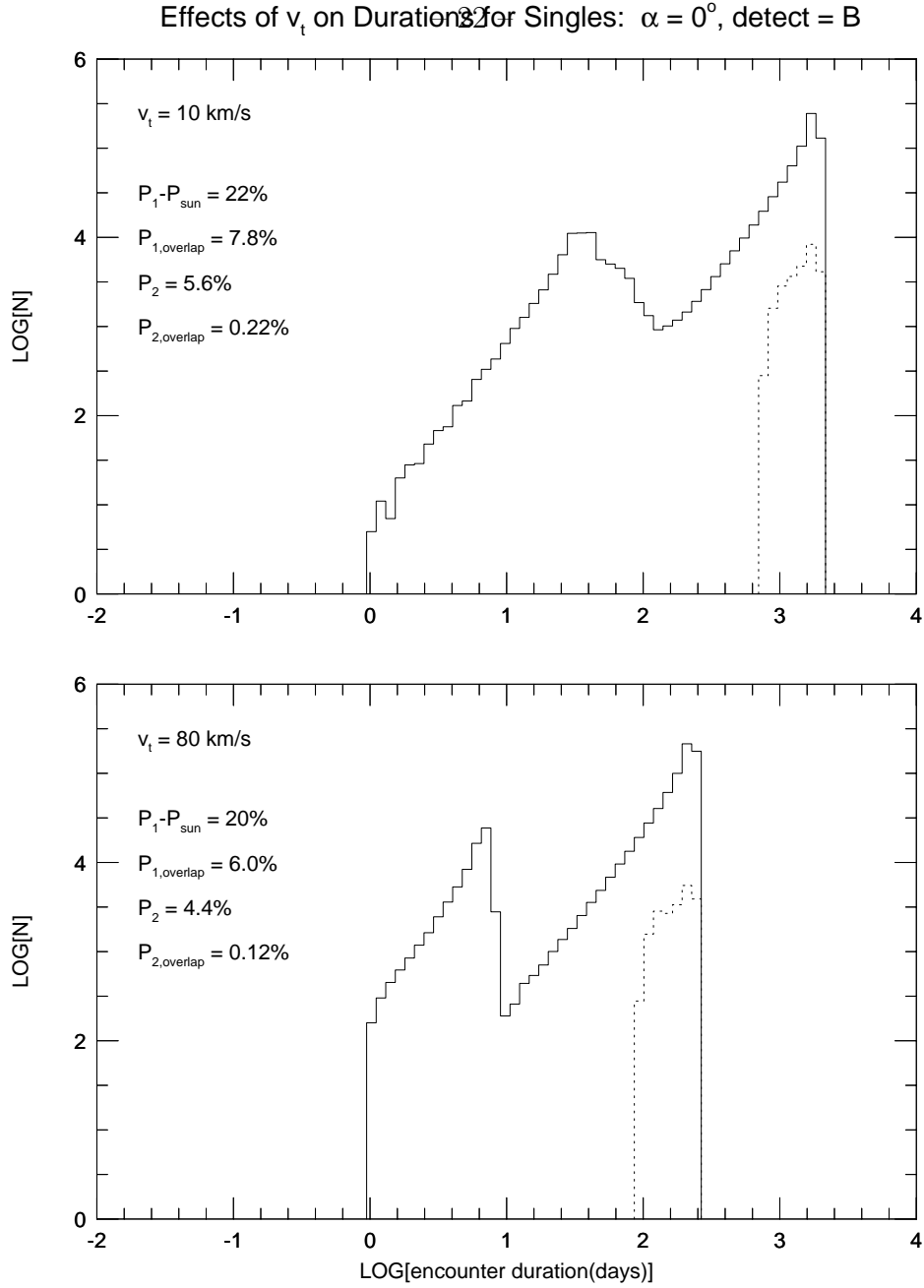}
\vspace{-0.5 true in}
\caption{Distribution of encounter durations for isolated \ev s. 
Detection criteria B are used to determine whether each \ev\ is
detectable.  
The duration is defined to be the time interval during which the projected
separation between the source track and the lens is smaller than $2\, R_{E,i}$,
where $R_{E,i}$ is the Einstein radius of the lens.  
The \ps\ is a power-of-$2$ 
system with five Jupiter-mass planets; the orbital plane was coincident 
with the lens plane ($\alpha=0$).  Top panel: $v_t = 10$ km/s.
Bottom panel $v_t = 80$ km/s.  
The dashed curves show the duration of overlap \ev s. 
}
\end{figure}

We used a truncated (5-\pl) power-of-2 system to explore the influence
of changing $v_t$ on the rate of events and characteristics of \ev s.
\footnote{The reason for the truncation was simply to achieve 
better sampling, i.e., a higher density of source tracks.} 
In the simulations whose results are shown in Figure 2, we took 
$\alpha=0;$ i.e., the orbital plane coincided with the lens plane.    
Holding $\alpha$ fixed, we systematically varied $v_t.$ 
The distribution of \ec\ \du s for $v_t=10$ and $80$ km/s are shown in Figure 2.
A comparison between the two cases clearly illustrates that, in keeping 
with the predictions,
the 
overall rate of detected \ev s is larger for all types of \ev s when
$v_t$ is smaller. This effect is most pronounced for repeating events
which involve several planets, since the probability of a repeating event
behaves like a product of detection probabilities for each planet separately.
The rate (per year) $P_{2,overlap}$ of overlap doubles demonstrates this well,
since an overlap double involves at least three objects, two of which must be
planets.  $P_{2,overlap}$ is larger for $v_t = 10$ km/s
by a factor of $\sim 1.8$. 
When $v_t$ is comparable to the orbital velocity, as it is in the
top panel, there is a clear dispersion in the distribution 
of encounter durations of short-duration \ec s in which
a \pl\ serves as a lens.  This dispersion is due to the influence of the 
\pl s' orbital velocities; encounters can be lengthened or shortened depending
on the angle between the planet's orbital motion and the transverse source
velocity.
In contrast, for $v_t = 80$\  km/s,
the peak due to planets is sharp and well-defined.  

As $\alpha$ increases, the projection of the orbits onto the lens plane
become ellipses. The projected orbital speed along the
semi-major axis is the same as before, but the projected orbital speed 
along the transverse direction is smaller. Thus, along some directions of
approach the effects associated with the finite size of $v_t$ become 
less pronounced.

\subsection{Systematic variation of orbital inclination}

\begin{figure} 
\vspace{-1 true in}
\plotone{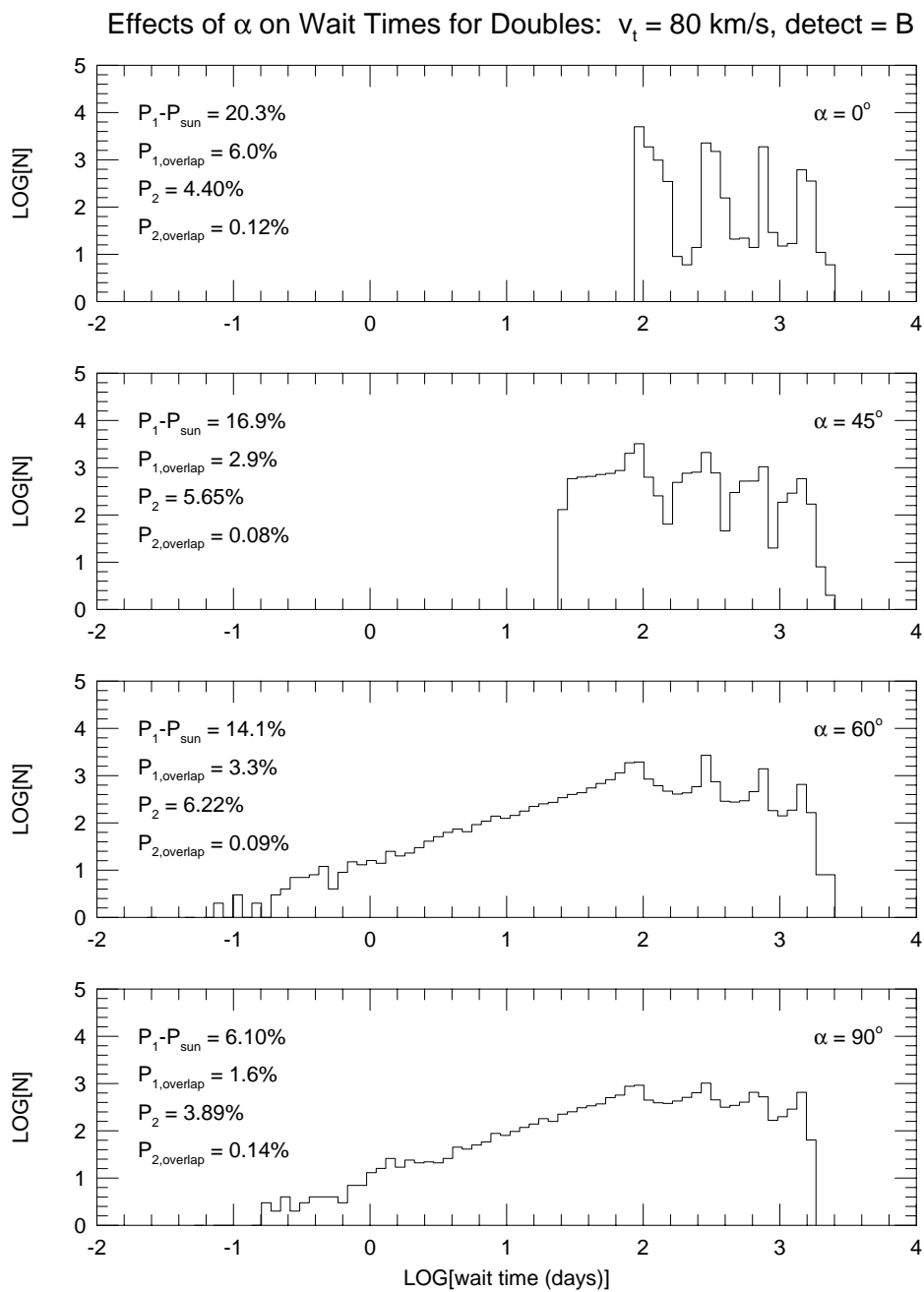}
\vspace{-0.5 true in}
\caption
{Distributions of durations of, and
wait-times between, encounters for events with one
repetition.  The model \ps\ serving as a lens is the 5-planet power-of-$2$
system.}
  
\end{figure}

To systematically
 test the results of changing $\alpha,$ we again used the 5-planet
power-of-$2$  system, this time keeping $v_t$ fixed at $80$ km/s.
The results for $4$ 
simulations ($\alpha=0^\circ, 45^\circ, 75^\circ,$ and $90^\circ$)
are shown in Figure 3.
For $\alpha = 0^\circ$, four peaks in the distribution of wait times
between \ec s are sharp and
clearly visible; these correspond to wait times between encounters with the
central star and encounters with one of the four outermost planets.  (The
innermost planet is so close that it serves as a lens only in overlap events.)  The
even spacing between the peaks is a signature of the power-of-2 model,
becoming about $0.3 \approx \log_{10} 2$ for planets with $a_i >> R_{E}$.
As the inclination increases, the dispersion mentioned in \S 3.         
appears; note that increasing inclination can only decrease wait
times, so that each peak is ``smeared'' out to the left until at high
inclinations a wedge-like shape is achieved.
The detection rates for non-repeating
events decreased with increasing inclination, but the relative 
detection rates for
repeating events increased, as predicted.
For $\alpha = 45^\circ$, the innermost planet's orbit comes within
$0.7 R_{E}$ of the star, placing it in the \zres\ most of the time.
The detection rates for doubles and overlap singles decrease
dramatically because of this, but they rise again at $\alpha = 60^\circ$ as
the second planet's projected orbit becomes small enough for overlap
\ev s to occur.  
At $\alpha = 90^\circ$ (edge-on), the motion of the \pl s 
brings all of them into the \zres, or even closer, part of the time.
This 
decreases overall detection rates for \pl s
in \wo s. The detection rates for
doubles decrease dramatically; this is because all planets lie along the same
line, and events which might have been doubles actually become triples,
quadruples, or higher-order events.
In fact, we found that for   
$\alpha = 90^\circ$, the rate (per year) $P_3$, $P_4$, and $P_5$ were all approximately
0.3\%. 
Thus, as mentioned in \S 3, 
a relatively large fraction of \ev s can exhibit multiple repetitions,  
as the track of the source sweeps across the ecliptic, crossing through
the \lr s of several \pl s and the central star.

\subsubsection{The Effects of Changing the Detectability Criteria}
\tightenlines

\begin{deluxetable}{lllllll}
\scriptsize
\tablecaption{Simulation Results:
 wide-orbit events for known and model systems}
\tablehead{\colhead{Detect \tablenotemark{(1)} } &
   \colhead{$P_1-P_\odot$ \tablenotemark{(2)} } &
   \colhead{$P_1^{overlap}$ \tablenotemark{(3)} } &
   \colhead{$P_2$ \tablenotemark{(4)} } &
   \colhead{$P_2^{overlap}$ \tablenotemark{(5)} } &
   \colhead{$P_3$ \tablenotemark{(6)} } &
   \colhead{$P_3^{overlap}$ \tablenotemark{(7)} } }
\startdata
\multicolumn{7}{l}{\bf Solar system, $V = 150$ km/s,
   $\alpha = 0^\circ$:} \\ \\
A & 0.3 & 2.2 & 0.5  & 0.01 & 0.0      & 0.0 \\
B & 1.8 & 6.0 & 1.4  & 0.04 & $3.0 \times 10^{-4}$   & 0.0 \\
C & 4.4 & 6.0 & 1.7  & 0.05 & $2.4 \times 10^{-3}$   & 0.0 \\
\hline
\multicolumn{7}{l}{\bf Solar system, $V = 150$ km/s,
   $\alpha = 75^\circ$:} \\ \\
A & 0.1 & 1.7 & 0.8 & 0.02  & $1.8 \times 10^{-3}$
   & $1.5 \times 10^{-4}$ \\    
B & 0.3 & 2.4 & 1.7  & 0.04  & $4.2 \times 10^{-3}$
   & $1.5 \times 10^{-4}$ \\
C & 1.5 & 2.4 & 2.3  & 0.04  & 0.01
   & $1.5 \times 10^{-4}$ \\
\hline
\multicolumn{7}{l}{\bf Solar system, $V =$ Gaussian,
   $\alpha = $ uniform:} \\ \\
A & 0.3 & 2.1 & 0.7  & 0.02  & $1.4 \times 10^{-3}$ & 0.0 \\
B & 1.4 & 4.2 & 1.8  & 0.04  & $3.2 \times 10^{-3}$ & 0.0 \\
C & 3.1 & 4.2 & 2.1  & 0.05  & $5.5 \times 10^{-3}$
   & $1.5 \times 10^{-4}$ \\
\hline
\multicolumn{7}{l}{\bf Power-of-2 system, $V =$ Gaussian,
   $\alpha = $ uniform:} \\ \\
A & 23.7 & 2.2 & 3.0  & 0.03  & 0.03   & 0.0 \\
B & 53.8 & 4.0 & 6.6  & 0.08  & 0.06   & 0.0 \\
C & 55.9 & 4.0 & 6.6  & 0.08  & 0.08   & 0.0 \\
\hline
\multicolumn{7}{l}{\bf Power-of-3 system, $V =$ Gaussian,
   $\alpha = $ uniform:} \\ \\
A & 13.0 & 1.9 & 1.5  & 0.03  & 0.01   & 0.0 \\
B & 29.3 & 3.6 & 3.1  & 0.06  & 0.02   & 0.0 \\
C & 30.4 & 3.6 & 3.1  & 0.06  & 0.02   & 0.0 \\ 
\enddata
\tablenotetext{(1)}{Descriptions of the detection criteria can be found in
the text.  All probabilities are given as the rate (per year); these
correspond to the percentages of the number of events
in which the central star 
was the only lens encountered, and in which the magnification
reached at least $A_{min} = 1.34$.  
}
\tablenotetext{(2)}{Rate (per year) of isolated (non-repeating) 
events (one peak in
the light curve). In this column we include only \ev s 
in which a single \pl-lens was \ec ed.} 
\tablenotetext{(3)}{Rate (per year) of non-repeating events 
which exhibited
evidence of lensing by two masses. In this column we include only
 overlap \ev s;
in these cases the two lenses were almost 
always the central star and the innermost
\pl .}
\tablenotetext{(4)}{Rate (per year) of events with one repetition
 These are not overlap \ev s; two well-separated masses were
\ec ed. In most cases these two masses are the central star and
the second \pl\ out, but there are other contributions as well.}
\tablenotetext{(5)}{Rate (per year) of single-repetition 
events in which one component
consisted of overlapping \ec s; 
the repetition was due to lensing by a third mass.}
\tablenotetext{(6)}{Rate (per year) of events with two repetitions;
 all lenses were well separated, with no overlap.}
\tablenotetext{(7)}{Rate (per year) of double-repetition events
in which one component
showed evidence of 
overlapping \ec s; the two repetitions were due to lensing by two other
well-separated masses. If the linear density of source tracks passing through
the power-of-2 and power-of-3 systems had been the same as for the 
Solar System, there would have been \ev s in these categories.}
\end{deluxetable}

To better learn how to 
optimize the returns from the \ml\ observations, we tested the effects
of varying the detectability criteria, from the most conservative set
of criteria (set A), to the most inclusive (set C). The results
are shown in Table 2, and in Figures 4--6. 
 
The planetary systems serving as lenses are  
our Solar System, the twelve-planet
power-of-2 system, and the $7$-\pl\ power-of-3 system.  
Figures 4 and 6 show the
duration and wait time distributions for events with one repetition
when using detection criteria A and C, respectively.
Figure 5 shows the cumulative distributions for the graphs plotted in
Figure 4. (Note that we have also kept track of the isolated \ev s 
that would have been detected in these simulations. The results for
isolated \ev s are summarized, together with those for \rpe s, in
Table 2 and are illustrated in graphs included in the companion paper [\rd\
\& Scalzo 1998].)

The detection 
criteria of set A are the most restrictive.  When our Solar System
serves as a lens, with
$v_t$ = 150 km/s, only Jupiter and sometimes Saturn are able to produce events
sustaining a magnification of $A_{min} = 1.34$ for longer than 1 day.  Any
other planets would be detected only by the follow-up teams, i.e., only if a
larger object (almost always the Sun) was encountered first.  Thus, 
wide-orbit 
planets slightly less massive than Saturn are difficult to detect
using a strategy based on a set of criteria similar to that of set A.  
The power-of-2 and power-of-3
models we examined contained only Jupiter-mass planets, so each planet was
detectable,
except for the small fraction of \ec s in which the source track just
grazes the \lr\ of the \pl . 
 We note, however,
that the criteria of set A are not powerful tools for the
discovery of solar systems  like
our own.

Using the detection criteria of set B (detecting even a first \ec\ 
at $A = 1.06,$ instead of 
$A = 1.34$)
generally more than doubled the detection frequencies in our most
realistic simulations, in which $v_t$ was chosen from a
Gaussian distribution and $\alpha$
was chosen from a uniform distribution.  
Planets could be detected at a lower peak magnification, and more
planets in the system were able to sustain a magnification of $A = 1.06$ for
the 1-day minimum duration required of the first encounter.  For the solar
system, Saturn was often detected as an isolated lens, or as the
first lens \ec ed, and Uranus and Neptune made occasional
appearances as isolated lenses.  
Note that, taken by itself, the effect of decreasing $A_{min}$
to $1.06$ should increase the rate by a factor of $2$ (instead of $4$);  
this is because, 
even with the set of detection criteria A, $n$ was
already equal to $2$ for the second \ec . 
Deviations from the factor of $2$ increase
 are mostly associated with \ev s which include
\ec s with a low-mass \pl, and are therefore seen primarily in the 
Solar System; these deviations are due to our use of a minimum
duration for the  detection of the first \ec .

Finally, the detection
criteria of set C (removing the requirement that the first encounter last
for at least a day) made Uranus and Neptune more regularly
 detectable as isolated lenses,
whereas before they were primarily detected as part of repeating events 
involving a larger mass.  These planets produce significant new structure in
the duration distribution of isolated \ev s (\rd\ \& Scalzo 1998),
as well as filling out the structures found the distributions of \rpe\ 
characteristics.  Even Pluto produces its own peak at
$10^{-2}$ days ($\sim 10$ minutes), although this is too short to be
detected in practice.  

To assess the relative benefits of the three detection strategies used,
we note that the largest increase in the rates of detectable planet-lens 
\ev s was realized by switching from the set of criteria A, in which $A_{min}$
for the first \ec\ is $1.34$ to set B, in which $A_{min}$
for both \ec s is $1.06.$    
The gains in switching from B to C, which eliminated the 
requirement that the first detectable \ec\ have a \du\ of 1-day,
could be substantial only if, like our own Solar System, the lens
\ps\ contains wide-orbit \pl s less massive than
Jupiter.

In general, the smaller the value of 
$A_{min},$ the longer the duration of the observed \ev\ anyway.
Thus, while frequent \mo\ is desirable, and more frequent \mo\ than is
presently achieved by the \mo\ teams would be very valuable, 
the key issue is sensitivity to \ec s in which the peak \mage\ may
be smaller than $1.34.$ We note that this result is likely to remain
valid, even when other effects are considered. For example, 
both blending and \fsse\ decrease the peak \mage . 
Thus, detection rates for \ev s subject to these effects 
may also be improved if \ec s which lead to lower \mage\ peaks can
be detected.

\begin{figure}
\vspace{-1 true in}
\plotone{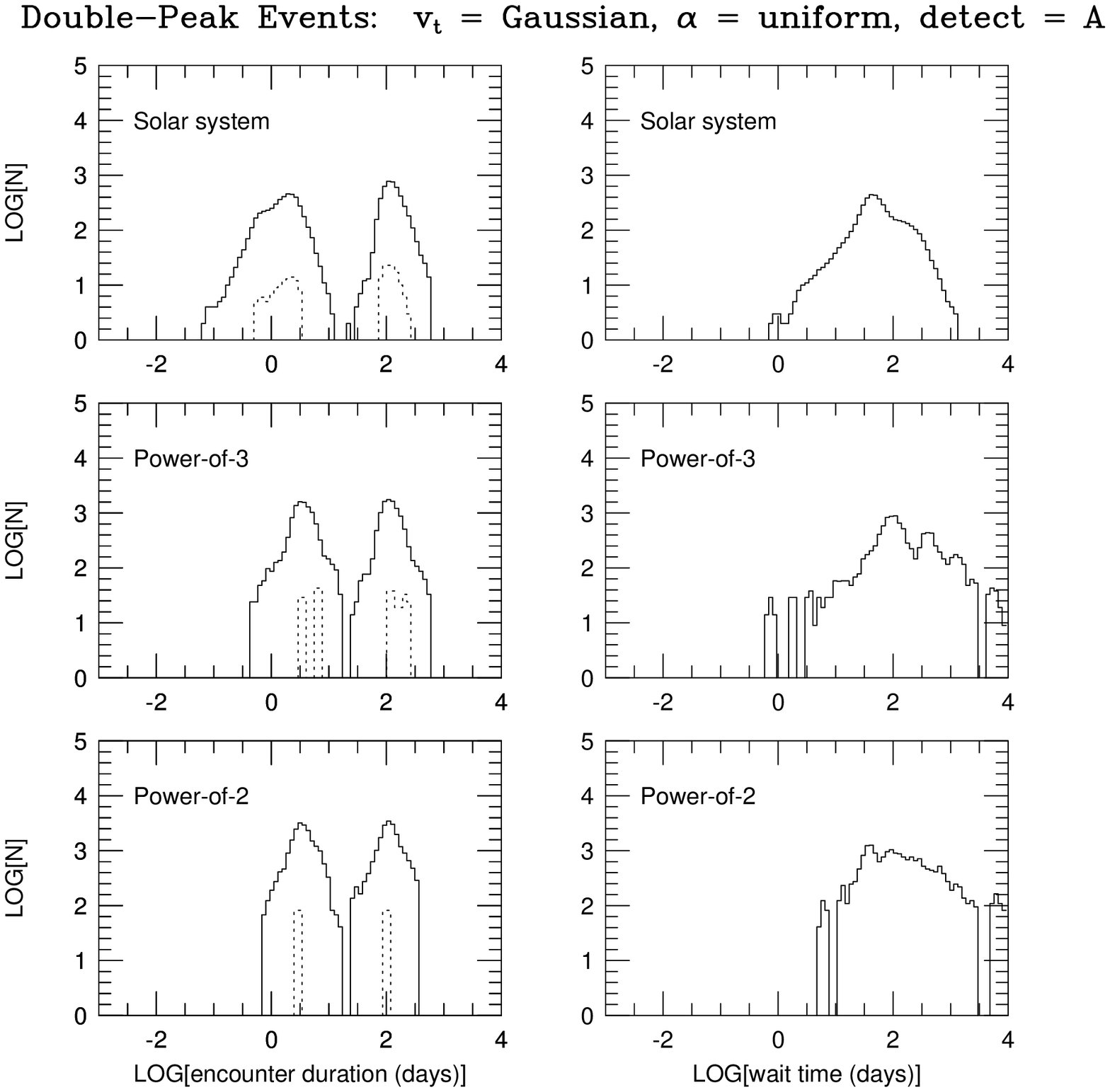}
\vspace{-2 true in}
\caption{Distributions of \ec\ durations and wait times 
between \ec s for events with one
repetition, using the detection criteria of set A.  
{\bf Durations:}\  For the Solar System (top left),
Saturn dominates the left-most peak, with Uranus and Neptune
appearing as a shoulder
to the left, at approximately 0.7 days.  Overlap doubles are
generally \ev s in which there is one anomalous-looking deviation
from baseline corresponding to an overlap \ev\ in which both Jupiter
and the Sun serve as lenses, and also a separate \shdn\ deviation
due to an independent \ec\ with one
of the outer \pl s.  
Overlap doubles are shown with a
dotted line:  overlap \ec s involving both Jupiter and the Sun 
are represented in the
right-most peak; the other \ec\ that occurred as part of these same
repeating \ev s was with either Saturn, Uranus, or Neptune, and  
produced the peak on the left.
{\bf Wait times:}\ For the Solar System, the predominant peak
is produced by doubles involving the Sun and Saturn; Uranus and Neptune
produce a broad shoulder to the right, at around $10^2$ days.}
\end{figure}

\begin{figure}
\vspace{-1 true in}
\plotone{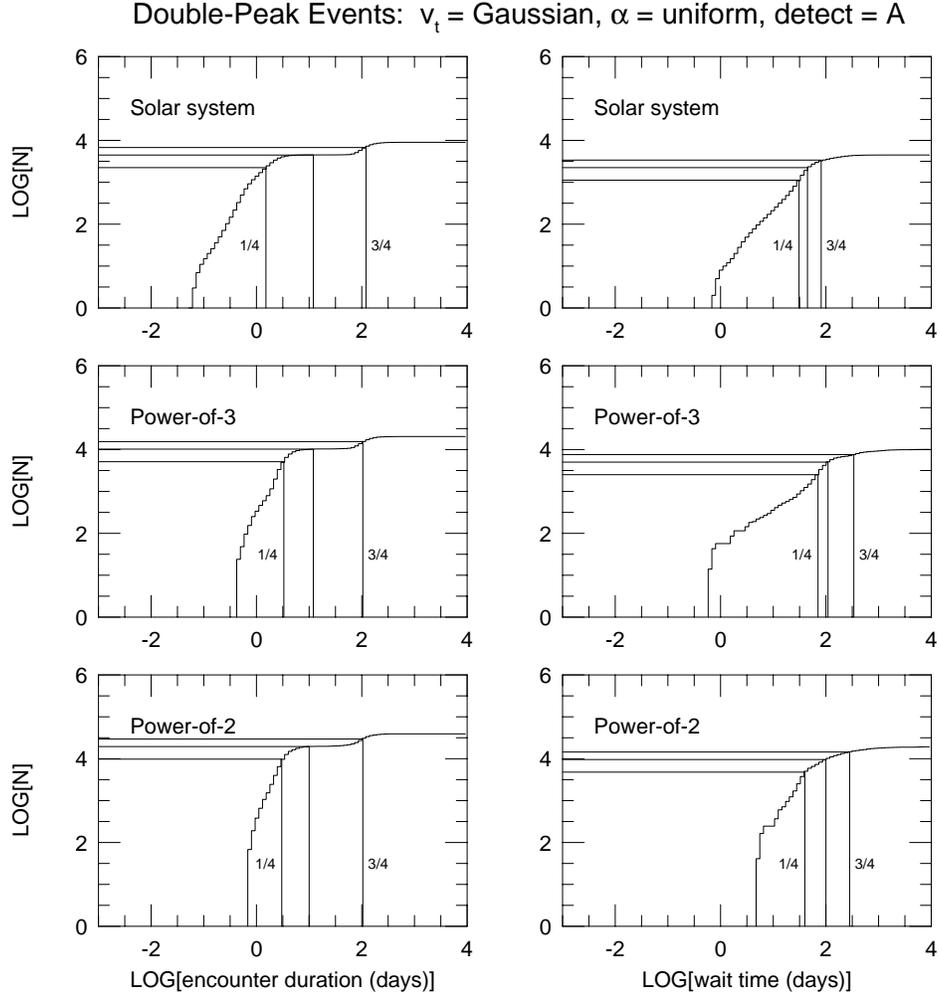}
\vspace{-2 true in}
\caption{Each panel shows the integrated area under the distribution to which
it corresponds in the previous figure.  
Encounter durations are shown on the left; wait times between
encounters are shown on the right.  Roughly $75\%$ 
of all encounters, and half of
the gaps between encounters ($75\%$ for the Solar System), last less than 100
days.}
\end{figure}

\begin{figure}
\vspace{-1 true in}
\plotone{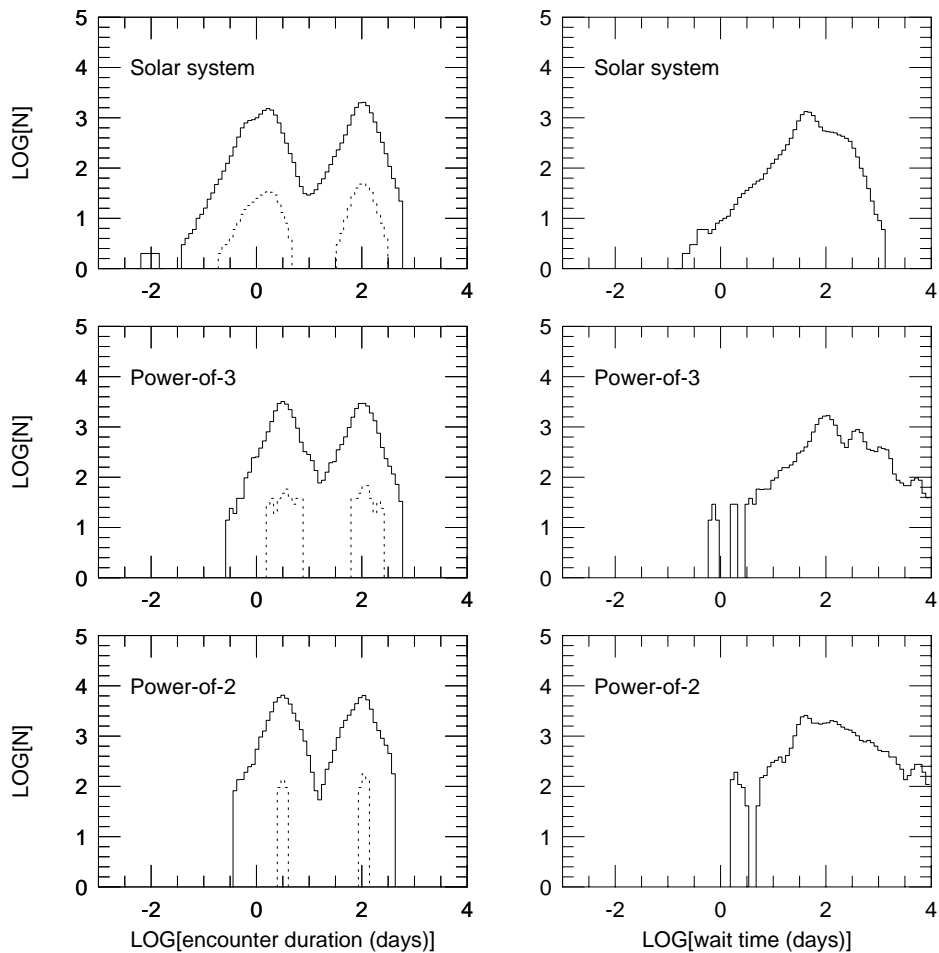}
\vspace{-2 true in}
\caption{Distributions of durations of and wait times between events with one
repetition, using the detection criteria of set C.  
Most of the structure visible in Figure 4, where
the detection criteria of set A were used, can also be seen here.
The overall event rates have
increased, and the gaps in the centers of the distributions have filled in.
A new peak corresponding to Pluto has appeared in the duration distribution
for the Solar System (top left); these \ev s are so short that, although we
may occasionally catch one magnified point on their \lc s,
 we are unlikely
to be able to follow them. }
\end{figure}

\subsubsection{General Features}

Perhaps the most interesting feature of the results is that 
they clearly indicate the feasibility of the search for \ps s 
containing \pl s in \wo s. 
For the most optimistic model, the 12-\pl\ power-of-$2$ model,
even the strictest detection criteria
yielded a rate 
 of  $4.5\%$
 for \ev s showing some
evidence of two or more objects in the system and $23.5\%$  
for isolated short-duration \ev s. 
\footnote{These rates are computed using the normalization
described in \S 2. That is, they are $100$ times the number
of \ev s of the type described, divided by the number of \ev s in which
the central star served as the lens in an isolated \ev, with 
peak \mage\ of the stellar \ev\ achieving a value greater than $1.34.$} 
For the detection criteria of set C, $10\%$ was the rate of \ev s showing some
evidence of two or more objects in the system, and $\sim 56\%$
was the rate of isolated short-duration \ev s.  
It is clear that, if the power-of-2 model is 
realized in nature with any great
frequency, we will be able to observe many \pl-lens \ev s even
during the next few years. Conversely, an absence of large numbers of
interesting \pl-lens \ev s, particularly if we use the detection criteria
of set C, would allow us to definitively falsify the hypothesis that most
stars are accompanied by power-of-$2$ \ps s. 

While 
it may not be surprising that such a radical model is verifiable
or falsifiable, the relative ease with which \ps s such as our own
can be discovered is indeed worthy of note. The rates of \wo\ \ev s   
range from $\sim 3\%$ for the detection
criteria of set A to $\sim 10\%$ for the detection
 criteria of set C.
This means that, even over the short-term, observing strategies
such as the ones we propose in \S 5, will lead to significant
(positive or null) results.
In \S 6 we discuss the likely results for \ps s in general. 

An important question for the observing teams is how long they must
continue frequent \mo\ in order to have a good chance of detecting
a repetition. Figure 5 indicates that 
$1/2-3/4$ of all \ev s with one repetition
will begin the repetition within $100$ days of the time the 
magnification associated with the first \ec\ dips below $1.06$.
In addition, 
a large fraction ($\sim 75\%$) of  overlap \ev s may be identifiable.  
In our simulations, overlap \ev s form a significant
subset ($1-4\%$) of all \ev s for all of the models (Solar System
through to power-of-$2$) and for all detection criteria (sets A through C).
This is simply due to the fact that each model
system has a \pl\ located just outside the zone for
resonant lensing. Note that, if overlap \ev s are not seen
at roughly this level, there are likely to be few resonant \ev s,
since the \pl s that serve as lenses in overlap \ev s are the 
ones that can be brought into the zone for resonant lensing 
as the orientation  of the \ps\ changes. 

\subsection{Finite-Source-Size Effects}     

\begin{figure}
\vspace {-1 true in}
\plotone{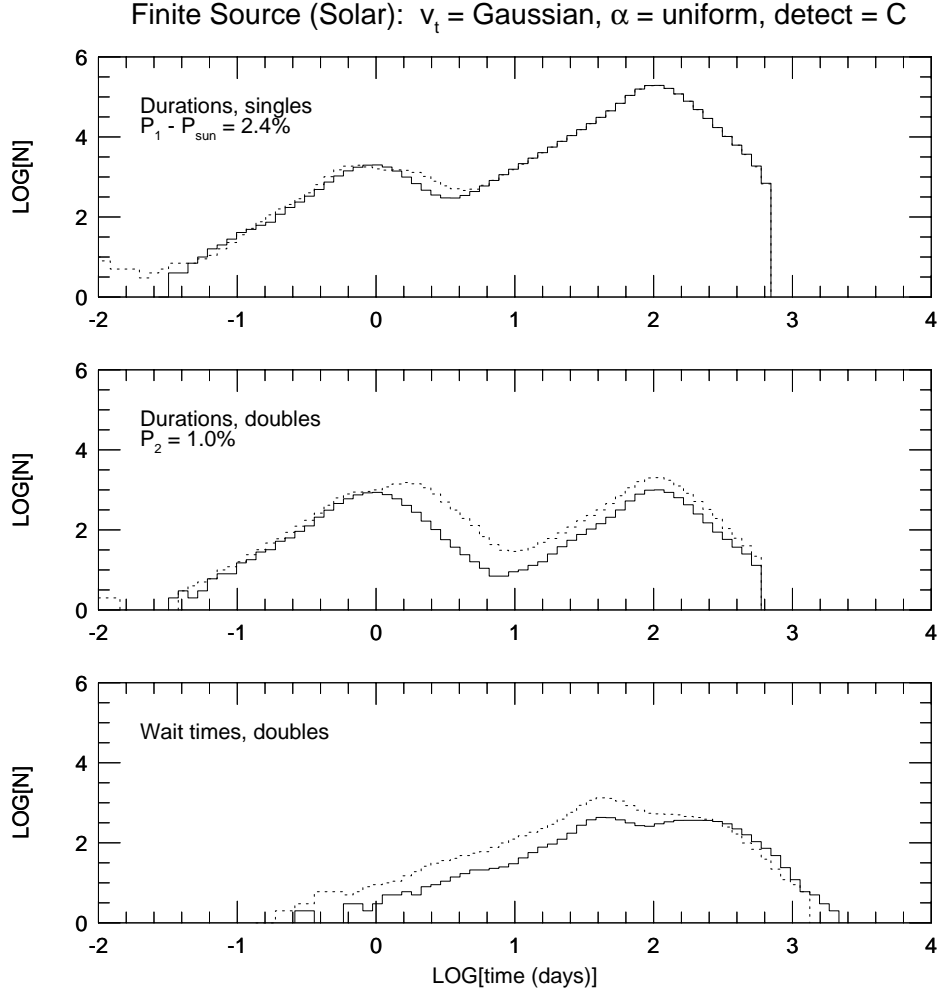}
\vspace {-2 true in}
\caption{Results of a 
Monte Carlo simulation 
that includes finite-source-size effects for a lensing system with
9 Earth-mass planets whose separations from the
solar-mass central star are the same as
those of the planets in our Solar System.  
We have assumed that the source radii are distributed  
uniformly from $3\, R_E$ to $10\, R_E.$
In addition, we assume that lensing by Earth-mass \pl s cannot be detected if
$R_S > 7\, R_E.$ Thus, roughly $43\%$ of the sources could not produce
identifiable lensing \ev s. For the remainder of the sources, we used
the approximation that $D_{min}\sim R_S.$ (See Figure 2 of the companion
paper [\rd\ \& Scalzo 1998].)
In these simulations we used: $v_t$ = Gaussian, $\alpha$ = uniform, detect = C.
For comparison,   
the distributions 
for the equivalent simulation of
lensing by the model of the  Solar System used earlier,
with \pl\ masses the same as those of our own Solar System,
and the sources taken to be point-like,
 are shown by the dotted curves.
}
\end{figure}

To better quantify the influence of finite source size, 
we have carried out simulations using the Solar System model
and a power-of-3 model to study the detectability of Earth-mass
\pl s. 
The placement of the \pl s was exactly as
in our previous simulations of the Solar System and the power-of-$3$
model, but the mass of every \pl\ was the same as that of the Earth.
We assumed that the source radii are distributed
uniformly from $3\, R_E$ to $10\, R_E.$
In addition, we assume that lensing by Earth-mass \pl s cannot be detected if
$R_S > 7\, R_E.$ Thus, roughly $43\%$ of the sources could not produce
identifiable lensing \ev s. For the remainder of the sources, we used
the approximation that $D_{min}\sim R_S.$ (See Figure 2 of the companion
paper [\rd\ \& Scalzo 1998b].) 
Note that, because the model \ps s in these simulations are
in the Bulge, and are lensing more distant Bulge stars, the \re\
of an Earth-mass \pl\ is approximately $1\, R_\odot.$

Thus, because we have assumed $R_S > 3\, R_E \sim 3\, R_\odot,$
we are not considering the contributions of the lensing of dim stars;
because such stars are likely to be numerous, their contributions could be
important, particularly when difference techniques are used to analyze the data.
This is because such techniques allow the lensing of stars not 
bright enought to
appear on a template to be detected. Furthermore, our assumption that
 $R_S < 10\, R_E \sim 10\, R_\odot,$ means that we are not considering
the contribution of giants. In fact, giants cannot contribute to
the lensing by Earth-mass \pl s, since their large radii can allow only
very small values of the peak \mage\ to be achieved. Their contributions
can be important, however, for the detection of more massive planets. 
     
We used $v_t$ = Gaussian, $\alpha$ = uniform, detect = C.
These results are summarized in Table 3 and are illustrated
for the Solar System model in Figure 7. 

Because their Einstein rings are smaller, Earth-mass \pl s are expected
to yield a smaller \ev\ rate. The rate would be 
smaller than that due to Jupiter-mass \pl s by a factor of
roughly $18$. Yet, in the power-of-3 model, where all of the \pl s in our
original model had a mass equal to that of Jupiter, 
the computed attenuation of the
Earth-mass system relative to the 
Jupiter-mass system is only a factor of $\sim 3.$  Even though
\fsse\ made \ev\ detection
impossible when the Earth-mass \pl s lensed roughly $43\%$ of the 
sources, the fact that the detection rate was higher for the remainder
of the sources provided a net increase in the detection rate over what might
otherwise have been expected. The relative increase is even more
pronounced for the solar-system model, since Saturn, Uranus, Neptune, and
Pluto are responsible for many of the isolated \ev s when a model
of the Solar System  serves as a lens, and these each
have a smaller Einstein radius 
 than Jupiter.

It is also interesting to make the comparison to the lensing
by this mock Solar System, with only Earth-mass \pl s,
when the sources it lenses are all point-like. We find that
finite-source size effects are associated with a significant
overall increase in the detection probability. 
Single-perturbation \ev s (top panel of Figure 7) would occur at a rate
of less than $0.5\%,$ were the sources point-like, whereas
doubles would occur at a rate of less than $0.2\%$. Thus,
finite-source size effects have increased the detection probability by a factor
of $\sim 5.$ (This, despite the fact that some sources were either too
small or too large to be detectably lensed in the finite-source-size case.)

Although more
realistic stellar surface brightness profiles and stellar population 
models should be used,
the basic result that Earth-mass \pl s are detectable should be robust.  
Also robust is the increased \ev\ \du\ for a range of values of
$R_S$. For a fixed value of $A_{min},$ this corresponds to
a larger value of the effective width, $w_i$, and therefore a larger
value of $n_i=w_i/R_{E.i}$.  
As predicted 
 and as demonstrated by the simulations, 
this can increase the detection probabilities
above what is expected if the source is point-like.

\section{Detection Strategies}

\subsection{Optimizing Our Ability to Detect Planets in Wide Orbits}

Three steps can improve our sensitivity to repeating \ev s.
(1) Improving the photometric sensitivity.
(2) Instituting reasonably frequent monitoring.  
(3) Being alert to repeats and to the odd-ball \lc s
associated with overlap \ev s. See Sackett (1997) for an
analysis of the strategies best suited to the discovery of
\pl s in the \zres .  

\subsubsection{Photometric Sensitivity} 
Photometric sensitivity  is the key issue for the detection of
\pl s in \wo s. (See also
the companion paper [\S 5, \rd\ \& Scalzo 1998].) 
The better the photometry,  the smaller the value of $A_{min}$ (the 
minimum peak \mage\ needed for reliable \ev\ detection), the 
larger the value of $n_i=w_i/R_{E,i}$ for each lens, and the  
larger the number of detectable \ev s. 
The increase is largest for repeating \ev s, since the
rate of such \ev s scales as $\prod_{i=1}^{k+1} n_i,$ where
$k$ is the number of repeats.

\subsubsection{Frequent Monitoring}

Not only are more \ev s discovered
when the  $n_i$ are greater than unity, but the \du\ of detected \ev s is also
longer, making less frequent \mo\ possible.
The optimal monitoring frequency therefore
depends on $A_{min}.$ 

\subsubsection{Oddball Events} 
The early folklore on repeating events said that they should be
discarded: the microlensing of light from a specific star is such
a rare event that, if a \ml-like signal is seen twice when monitoring the
same star, the repetition is almost surely a sign that the observed variation is
due to
something other than \ml .
Although it has been
pointed out that binary sources (Griest \& Hu 1992) and binary lenses (\rd\ \& Mao)  
can each lead to \ml\ \ev s that repeat, it is nevertheless the case that,
until recently,  
 one of the cuts used to eliminate
lensing candidates from further consideration is based on whether
the event appears to repeat. {\footnote {Of course the complication of
ensuring that a repetition is not due to stellar variability is serious,
since variability is far more common than \ml. 
Nevertheless, if a  star that has
never been observed to vary from baseline produces two deviations well-fit
by \ml\ models, it is certainly worth exploring the possibility
that the deviations are indeed due to \ml . In some cases blending or
finite-source-size effects may make it possible to carry out further
tests of the conjecture that multiple disturbances are due to \ml.}} 
   As a result, the \mo\ teams may not have systematically pursued the
implications of evidence of
repeating \ev s in their data sets.
The \ev s most likely to have been
missed are: (1) events that were not recognized in real-time, and for
which no alert was called. Software packages that sort through
the data {\it post facto} might have eliminated such \ev s from further
consideration. This is not a serious problem, because (a) it can be
corrected by updating the software if necessary, and (b) there do not
seem to have been many \ev s eliminated by this criterion (Axelrod 1997).
(2) events in which either \ec\ led to a peak with a small maximum
\mage . This may also not be a serious problem, unless the
low-\mage\ peak was so small that the photometric
sensitivity of the monitoring team does not allow it to be distinguished from baseline.
 In other cases, there would simply
be one \ec\ leading to an observable \ev\ and a second perturbation,
too small to be reliably identified as due to \ml . The frequency of
these situations can be determined {\it post facto}. 
(3) events in which both \ec s led to peaks with small peak
\mage . Neither perturbation would have been identified as a \ml\ 
\ev . As above, however, such \ev s can be found by 
sifting through the data with a software program designed to look for
them. It may, however,
 be difficult to reliably attribute them to \ml . 
The statistics of such \ev s can be interesting, however, because 
the numbers of those that
are due to \ml\ should be closely related to the numbers of other \rpe s
with larger peak \mage s and smaller distances of closest approach.   
(4) \ev s in which one or more deviation from baseline is short-lived.
Such \ev s present the most serious challenge to
\ev\ detection, because 
this description fits all \rpe s (except for \ov\ \ev s)
in which a \pl\ serves as a lens. This makes it clear that
\ov\ \ev s will be the easiest to detect at present. Repeats in
which the first rise from baseline is due to lensing by the \cs ,
will be the next easiest to detect, because their detection requires only that the
follow-up teams continue to carefully monitor a stellar-lens \ev\
which has apparently ceased. Repeats in
which the first rise from baseline is due to lensing by a \pl\ will be
the most difficult to detect with present strategies; the  changes
needed to detect such \ev s are the same as those needed to detect isolated
\ev s of \shdn , and are discussed in \rd\ \& Scalzo 1998.
   
In the companion paper we noted that the Bulge data collected by the
MACHO team already contains enough \shdn\ \ev s to demonstrate the monitoring
teams' capability to systematically detect such events, especially
with some modification in their observing strategy. Here we note that
repeating \ev s are also well within the realm of presently detectable 
\ev s. The first ``repeater" mentioned in the MACHO alert pages
is 96-BLG-4. The first \ec\ of that \ev\
was apparently a relatively low-magnification
Paczy\'nski \lc\ ($A_{peak}=1.9$)  of long duration ($\sim 111$ days),
and the second perturbation occurred in the autumn of 1997. There
is no information posted about the duration of \mage\ of the second
deviation from baseline.
The second ``repeater" is the pair 97-BLG-45 (which exhibited a peak
\mage\ of $4.8$ and a duration of $\sim 107$ days) and  97-BLG-47.
Although no information about the second deviation is provided, its
place in the sequence of 1997 Bulge events indicates that there was
likely not a long wait between the two \ec s, if indeed both
perturbations are due to \ml .
   
\subsection{Observing Programs To Discover Planets}

The design of the observing programs of the future should take the above
requirements seriously. There are likely to be several
 different ways to proceed effectively. For example, a system of \mo\
could achieve a detection rate $\sim 10$ times as high as the present system
does by (1) using photometry as good as or better than the present follow-up
teams, (2) using difference techniques to identify lensing of stars that are
otherwise (i.e., at baseline)
 below the detection limit, and (3) \mo\ frequently.
This would mean discovering ${\cal O}(1000)$ Bulge \ev s per year. Not only 
would the numbers of \ev s be high, allowing us to detect some very
low-probability \ev s (such as multiple repetitions), but the quality of the 
data would be so high that we would understand each \ev\ and the population
of lenses and sources much better than we do at present. 
Another approach might be to have a world-wide network of telescopes, each   
taking deep images of a few fields once or twice per night. The detection
rate per field and the quality of the data would be high, but the number of 
\ev s discovered per year would be smaller, since fewer stars are being 
monitored. 
The total number of \ev s discovered per year 
might be comparable to the discovery rate of the 
present MACHO team.

It is
worthwhile to consider what the teams using the wide-planet-search strategies
will learn, even if nature has been so unkind as to neglect to
provide most stars with \pl s. Planet-motivated
investigations will yield interesting fruit
regardless of the size of the population of \pl s. 
First, they will increase the detection rates of all \ev s,
particularly \shdn\ \ev s.
Second, frequent \mo\ with good photometry will
allow us to learn more about each \ev\ detected.
Common astronomical effects, such as stellar binarity,
blending, and \fss ,  
 are expected to significantly
affect the shape of lensing \lc s, 
introduce time dependence into the spectra, and even produce 
apparent repetitions.   
(See e.g, Griest \& Hu 1992; Mao \& Paczy\'nski 1991;
Mao \& \rd\ 1995;    
\rd\ \& Esin 1995; Kamionkowski 1995; 
Loeb \& Sasselov 1995; Simmons, Willis \& Newsam 1995; \rd\ \& Mao 1996;
Sasselov 1997;
\rd\ \& Perna 1997; \rd\ 1997) Studying the
manifestations of these effects in the data sets can
(1) can break the degeneracy of individual \lc s, and 
(2) allow us to learn more about the
populations of sources and lenses.
Along the direction to the Bulge, we will be able to learn 
a good deal about the stellar luminosity
and mass functions, and the binary fraction, as well as the distribution
of binary properties.  
The  information we collect can inform our design of the next
generation of \ml\ observations. For example, satellite projects have been
proposed and preliminary calculations indicate that they are likely   
to be productive. (See, e.g., Boutreux \& Gould 1996 and references therein.)
The detailed observations that would be made as part of the search
for \pl s in \wo s would provide  solid input, useful for the detailed
 planning needed for  
such space-based projects.

\subsection{Useful Modifications Of Existing Programs}

\subsubsection{The Monitoring Teams}

 The present strategy of the \mo\ teams
would allow them to reliably detect the first \ec\ of a \rpe\ only if the first
mass serving as a lens is the \cs , and if the peak \mage\ is large enough. 
There would be some chance 
of detecting the first \ec\ if the first
mass serving as a lens is a \pl\ with a mass comparable to or greater than
that of Jupiter.   
It is possible, however, that even the present data sets 
may contain evidence of repetitions in addition to those
listed on the alert web pages (http://darkstar.astro.washington.edu).
 If this is so, such evidence might
best be discovered through 
searching the existing data sets for repetitions, perhaps in
conjunction with a spike analysis.
Particularly in fields visited every day, a spike analysis can provide
valuable information. The ratio of spikes expected to be
part of a \rpe\ to those expected to be isolated depends on the
number of \pl s, and on the spatial and
mass distribution among the \pl s. 
It should mirror the overall ratio of \rpe s to isolated \ev s:
for systems similar to the Solar System this ratio might be on the order
of unity, while for power-of-$n$ type systems it could be as low
as $0.1$.

The present-day \mo\ teams may also be able to detect some repeating
\ev s when the first mass to serve as a lens is a \pl\ less
massive than Jupiter. But to do this reliably
 would require some modifications of
their observing strategy. Because the modifications likely to be
useful are exactly those that would allow the \mo\ teams to have
a better chance of detecting isolated \ev s of
\shdn, they are discussed in the companion paper (\rd\ \& Scalzo 1998).
 Here, we simply
mention the possibility of selecting a subset of their monitored fields for 
more frequent \mo .

\subsubsection{The Follow-Up Teams} 

The follow-up teams need to continue
to monitor all \ev s after the flux has apparently fallen back to
baseline. 
It should be possible for the follow up teams to
discover repetitions in most cases in which a star is the first lens
encountered. 
For the \ps s we have simulated,
roughly $1/2-3/4$ of all other repetitions could be detected if
monitoring were continued for $100$ days.
In addition, overlap \ev s would not require a significant wait-time,
since the second encounter would start (i.e., be associated with $A > 1.06$)
even  before the first encounter ended.

Note that continued \mo\ may be useful even 
when the first \ev\ observed is a resonant lensing  \ev , since 
(a) if the observed \ev\ was due to a \pl, there may
well be a second planet, and (b) if the observed \ev\ was due to a 
binary lens, one or
both members of the
 binary (or even the combination) may support a
\ps . 

We note that, even when the \mo\ teams have identified a \shdn\ \ev\ that 
appears to have ceased before more frequent \mo\ could begin, it is worthwhile
for the follow-up teams to continue to monitor the flux. If the
short-time-scale \ev\ was due to lensing by a \pl, the
result of frequent \mo\ 
could be the detection of a repetition due to a moon revolving about the \pl.
If a moon exists and if we were fortunate in the orientation of the
source track, then a repetition could occur and should be detected
within a matter of days.
A more delayed repetition could be due to later lensing by the
central star or by another \pl\ in the system.  If they can detect
\ev s with $A_{max}$ as small as $1.06$, the follow-up teams
 will be at least twice as likely to discover such \ev s as are the
\mo\ teams.
 
\section{Expectations}

What results are likely to be derived if the strategies sketched in
the previous section are utilized? It is difficult to answer this
question, because we know so little
about how common \ps s are, and about
the distributions of planetary masses and
orbital periods. Because, however, planetary systems have begun to be discovered
in the Sun's local neighborhood, it is beginning to seem
likely that many, perhaps most stars support \ps s.

If this is so, then the preliminary results represented by the
known \ps s
provide encouragement that \ml\ will play an important
role in the discovery of \pl s.
Systems with confirmed \pl s in orbits that would
be perceived by us to be either resonant or wide, were the system
to be placed in the Bulge, comprise roughly $1/4$
of all of the confirmed planetary and binary brown dwarf systems.
 PSR B1620-26 and the brown-dwarf system  Gl 229 would
produce larger numbers of isolated \ev s than our Solar System, but
fewer repeating \ev s. PSR 1257+12 would also lead to more isolated
than repeating \ev s.  
In addition, 55 Cnc and HD 29587 are
excellent candidates for lenses that would lead to resonant
\ev s. 
{47 Uma has an estimated
orbital separation slightly less than 0.8 $R_E$ for
$D = 10 kpc$, $x = 0.9$. It may also be a good candidate for detection.}
Thus, considering only confirmed systems, we find that  
both of the pulsars with \pl s have \w-o\ \pl s, 2 of the 11 brown
dwarf systems have either a resonant or \w-o\ \pl , and
2 of the 9 \ps s also have either a resonant or \w-o\ \pl .  
We note that, since \pl s in \wo s, and even those in the \zres ,
 are certain to
be under-represented in our present census of \pl s, inferences
based on the systems discovered so far may be conservative.
Thus, our present knowledge of \ps s makes it seem
likely that the observing teams
will observe some events   
over the next few years.

More important than the detection of any individual \ev, however,
is our ability to extract information about the population of \ps s
in the regions surveyed by the \ml\ teams. We would like to
know the answers to basic questions: what fraction of stars have
\ps s? What are typical numbers of \pl s in a single \ps ? 
What are the distributions of \pl\ masses and orbital periods? 

If the teams begin to discover \pl s,  
they will likely be able to 
establish the statistics of and distributions of
properties among \ps s in the Bulge.
\S 6.2 explores this issue
in more detail.

\subsection{The Relative Numbers of  ``Resonant" and ``Wide" Events}

Without knowing more about the distributions of planet properties--
which is exactly
what we are trying to learn about through the proposed \ml\ studies--
it is not possible to make definitive predictions for the relative
numbers of events due to \pl s in the zone for resonant lensing, 
and \ev s due to \pl s in \wo s.
We do know enough, however, to understand the issues that
determine the relative \ev\ rates.

If we place a Jupiter-mass planet in a resonant orbit, the chance of
detecting evidence of the planet's presence is close to $20\%$
The present observing set-up is optimized to discover this type of
\ev . If we place the same \pl\ in a wide orbit, the probability of
detecting an isolated \shdn\ \ev\ due to the \pl\ is smaller,
approximately $(3\, n)\%,$ 
and the probability of detecting a repeating \ev\ in which the central star
serves as the other lens is $(6\, n^2 \%)/(\pi\, a).$ 
Thus, with the present
observing strategy, individual examples of
each type of wide-orbit \ev\ are much less likely to
be observed than a resonant \ev. 
Even with the present set-up, however,
\pl s in \wo s may be detected at a rate comparable to 
or even larger than the rate of detecting \pl s in  
resonant orbits, simply because
on average there may be on the order of $10$ times as many of them as there 
are \pl s in resonant orbits. 

Two factors can enhance the relative 
probability of detecting \pl s in \wo s.
The first is that improved sensitivity to 
\shdn\ \ev s and   
better photometric sensitivity 
can significantly increase the detection
rates for all wide-orbit \ev s, particularly
of repeating \ev s.
The second is the influence of \fsse , which
can also  
effectively increase the  
value of $n$, even for a fixed value of $A_{min}.$

\begin{deluxetable}{lllll}
\scriptsize
\tablecaption{Detection Rates for Planets in Resonant and Wide Orbits:
\hfil\break
Point and Finite-Sized Sources}
\tablehead{\colhead{Detect \tablenotemark{(1)} } &
   \colhead{$P_{res}$ \tablenotemark{(2)} } &
   \colhead{$P_1-P_\odot$ \tablenotemark{(3)} } &
   \colhead{$P_1^{overlap}$ \tablenotemark{(4)} } &
   \colhead{$P_2$ \tablenotemark{(5)} }} 
\startdata
\multicolumn{5}{l}{\bf Solar system with point-like sources, $V =$ Gaussian,
   $\alpha = $ uniform:} \\ \\
A & 5 & 0.3 & 2.1 & 0.7 \\
B & 5 & 1.4  & 4.2 & 1.8  \\
C & 5 & 3.1  & 4.2 & 2.1  \\
\hline
\multicolumn{5}{l}{\bf Solar-like system w/finite sources, 
all \pl s of Earth mass, $V =$ Gaussian,
   $\alpha = $ uniform:} \\ \\
C & 0.3 & 2.4  & 0.7 & 1.0 \\
\hline
\multicolumn{5}{l}{\bf Power-of-3 system with point-like sources, $V =$ Gaussian,
   $\alpha = $ uniform:} \\ \\
A & 10-20 & 13 & 1.9 & 1.5 \\
B & 10-20 & 29 & 3.6 & 3.1 \\
C & 10-20 & 30 & 3.6 & 3.1 \\
\hline
\multicolumn{5}{l}{\bf Power-of-3 system w/finite source, 
all \pl s of Earth mass, $V =$ Gaussian,
   $\alpha = $ uniform:} \\ \\
C & 0.7-1.3 & 4.9 & 0.7 & 0.5 \\
\enddata
\tablenotetext{(1)}{Descriptions of the detection conditions can be found in
the text (\S 4.2).  All probabilities are given in terms of the rate (per year);
this corresponds to the percentage of the number of events
in which the central star
 was encountered by itself, and in which the magnification
reached at least $A_{min} = 1.34$.  
}  
\tablenotetext{(2)}{
We have 
have used the results of Bennett \& Rhie (1996) to compute the probability
that a \pl\ located in the \zres\ will be detected. 
They considered a mass ratio of $10^{-5}$, which, 
corresponds to an Earth-mass \pl\ if the \cs\ has a mass
of $\sim 0.3 M_\odot.$ Thus, when we use the results from
Bennett \& Rhie (1996) to derive the number of resonant \ev s
expected in our models, 
we may be making a slight overestimate.  
To be in line with $A_{min}$ for the wide-orbit calculations,
we have assumed that the observed resonant-\ev\ perturbation
should be a $6\%$ deviation from the underlying \lc , 
and have interpolated between
the results for the computation for $4\%$ and $10\%$ effects
(Bennett \$ Rhie 1996).
We have also averaged uniformly weighted
contributions for all of the cases they considered,
which correspond to $r\sim 1, 2, 4.3, 10,$ even though for the wide-orbit
case, where we also effectively performed a linear average, we
required $3<r<7.$
For the Solar System, we have assumed that there is a $25\%$ chance that one
of the \pl s is in the \zres; 
for the power-of-3 model we have assumed that there
is a $50-100\%$ chance that a \pl\  is in the \zres.}
\tablenotetext{(3)}{Rate (per year) of
isolated  (non-repeating) events 
with no overlap (see below).}
\tablenotetext{(4)}{Rate (per year) of isolated (non-repeating)
overlap events. In this case the two lenses were the central
star and the innermost \pl .
}
\tablenotetext{(5)}{Rate (per year) of events with one repetition. 
 Overlap \ev s are not included.}
\end{deluxetable}

Table 3 illustrates the situation for our Solar System, and for a Solar System
composed of Earth-mass \pl s. (See also Figure 7.)
The first three lines pertain to
a \ps\ identical  
to our own Solar System, placed in the Bulge.
Since, averaging over angles, Jupiter has a $20-25\%$ chance of being     
viewed in the \zres, and since there is a $\sim 20\%$ chance of
detecting a Jupiter-mass in the \zres,
we have estimated that
there is a $\sim 5\%$ chance of detecting evidence of the
Solar System through a resonant lensing \ev . For the detection
criteria of set A, the resonant-lensing signature
 would be the dominant mode of detection. Changing the
detection strategy to allow $A_{min}=1.06$ ($n=2$), allows wide-orbit
\ev s to dominate, with overlap and repeating \ev s having a combined
detection
rate of $6.3\%$ ($7.3\%$ if we eliminate the requirement of a
$1-$day \du\ for the first \ec ). Thus,
overlap and repeating \ev s are competitive with resonant lensing \ev s,
even if we assume that we may catch
most of the overlap \ev s but only $1/2$ of the repeating \ev s. 
Isolated \pl-lens \ev s would be found at a rate of $1.4\%$ and $3.1\%$
for detection criteria B and C, respectively.  These results indicate
that the
conjecture that all stars have \ps s similar to our Solar System,
predicts  
the detection of resonant, repeating, and \shdn\ isolated \ev s in roughly
equal numbers. If, however, all of the \pl s in the model Solar System 
were of 
Earth mass, the detection rate would fall. The fall is much more precipitous
for resonant \ev s, which are $5$ times less likely than the combined
rate of repeating and overlap \ev s and $8$ times less likely than
isolated \shdn\ \ev s.

For the power-of-3 model, we have assumed that the probability
of finding a \pl\ in the \zres\ ranges from $50\%$ to $100\%.$
The general pattern
of relative rates is similar
to that for the Solar System. For Jupiter-mass \pl s,
resonant events are competitive and
can even be dominant when the detection criteria of set A are used. 
The detection criteria
B and C increase 
the detection rate of overlap and repeating \ev s.
Isolated \shdn\ \ev s can also be important, and can occur at a rate
higher than  
resonant \ev s. 
Finite-source-size   
effects decrease the overall detection rates, making the 
repeating and overlap \ev s as common as resonant \ev s, 
and increasing the 
relative importance of isolated \pl-lens \ev s. 

The general pattern of relative rates illustrated by these examples is likely to 
be reflected in our data sets.
That is,  when \fsse\ are not important, the wide-orbit 
discovery channel we
have studied is competitive with the resonant-\ev\ channel. 
Repeating and overlap \ev s tend to occur at a lower rate than
either resonant or isolated \shdn\ \ev s, but the rates of the latter
two types of \ev s are comparable.  
When \fsse\
are important and/or when a strategy to optimize the discovery of \pl s
in \wo s is implemented, discovery of wide-orbit \pl s may dominate.
Isolated \shdn\ \ev s should be the \pl-lens \ev s most frequently detected,
and the combined rate of repeating and overlap \ev s may be 
comparable to or, depending on the influence of finite-source size,
 somewhat
larger than the rate of resonant-zone lensing \ev s.

\subsection{The Populations of Planetary Systems and Low-Mass MACHOs}

\subsubsection{Planetary Systems: General Considerations}

We found in \S 4 that we could hope to test simple hypotheses about specific
types of \ps s, even within the next few years. For example, do most
stars have \ps s similar to the Solar System? or similar to the power-of-2
or power-of-3 models?    
Of course it is most likely that \ps s come in several different varieties.
We can use the data to systematically extract information
about lenses with a possibly complicated distribution of \ps\
properties as follows.

Repeating and resonant events each allow us to determine specific
features of the \ps\ that served as a lens. Each provides the value
of the projected separation between one planet and another object
(either the sun for resonant events [and most repeating \ev s],
or a second planet for some repeating \ev s) in the \ps. Each allows
the derivation of a mass ratio.
\footnote{
The extraction of \pl\ parameters, the degeneracies in those
parameters, and the degeneracies between the \pl-lens interpretation and other
effects have been studied for \pl s in the \zres\  
by Gaudi \& Gould (1997) and Gaudi(1997).
Parameter extraction should generally be more straightforward for
\pl-lenses in \wo s. If the lens separations are very large, then the
\lc s are much like the standard point-lens \lc s; if the separations
are smaller, the fit given by \rd\ \& Mao (1996) can be applied.
For \rpe s, it may be possible to derive some constraints from the conditions
that both \ec s were due to lensing of one star of fixed radius, and
that the light contributed by the lens system was essentially the same for
both \ec s.} 
 If more information is available,
derived from evidence of blending or finite-source-size effects, for
example, it can even be possible to place reasonably tight limits
on the spectral type of the \cs\ and/or the mass of the planet-lens
(\rd\ 1998a,b).
If we are able, therefore, to discover and analyze a number
of repeating and/or resonant planetary-system-lens \ev s, we will, by
also including the influence of observational selection effects, be
able to derive some of the characteristics of the population
of \ps s in the Galactic Bulge and elsewhere. 

Consider 
a population of \ps s that act as lenses. Let 
${\cal P}_{res}(m)$ be the probability that a \pl\ of mass $m$ is
in the \zres, and ${\cal P}_{w}(m,a)$ be the probability that 
a \pl\ of mass $m$ is in a \wo, with separation $a$ from the central star.
As described above, discovering and measuring the rates
 of repeating and resonant \ev s 
constrains the form of these probability functions.
Furthermore, there is a consistency check, since
the values of ${\cal P}_{res}(m)$ and ${\cal P}_{w}(m,a_w)$
must be mutually consistent. In particular there is a close relationship between
the rate of resonant \ev s ands the rate of overlap \ev s. Among
wide-orbit \ev s , there is a similar close relationship between
the rate of overlap \ev s and the rate of \rpe s. Observations of 
resonant, overlap, and repeating events can help to constrain the
functional form of ${\cal P}_{w}(m,a_w).$  
The integral, $\int_{a_w}^\infty da\, {\cal P}_{w}(m,a),$
then predicts the rate of isolated \shdn\ \ev s that should be due to
lensing by \ps s. We note in addition, that the aggressive
and systematic study of blending and \fsse\ in all isolated
\ev s of \shdn\ can also be helpful. Indeed, such study can provide direct
confirmation that some such \ev s, those for
which the \cs\ of the \ps\ provides a significant
fraction of the baseline flux, are due to lensing by \pl s.

\subsubsection{Low-Mass MACHOs}

Should the predicted rates of detectable
short-duration \ev s be larger than the numbers actually observed,
then the analysis of \ps\ lenses must be
carefully reconsidered. We expect, however, that if there are
deviations from the predictions, they should be 
because the observed rate is higher than the predicted rate.
This is because \pl s in \ps s are not the only
lenses expected to produce short-time-scale \ev s.
For example, orbital dynamics may lead some \pl s
to be ejected from their home \ps .
If even $10\%$ of massive planets
are so-ejected, then, considering that the \ml\ observations
are sensitive to the planet debris of many generations of stars,
some of these ejected \pl s could be detected. In addition,
it is certainly possible that a significant fraction of Galactic
dark matter 
exists in the form of low-mass objects.
It is an interesting fact that, without the information  provided
by repeating and resonant \ev s, it will not be possible to
determine whether there is a low-mass MACHO component in directions
in which lensing by stellar systems contributes significantly to the
rate of \ml .
Thus, when stars contribute significantly to the rate of lensing,
 the detection of repeating and resonant \ev s is important 
in helping us to 
learn about any Galactic component of low-mass MACHOs.
 
In summary, a statistical analysis of the data on different types
of planet-lens \ev s can provide useful information. The 
planetary systems in each galaxy can be characterized 
by a  distribution function of physical planetary  separations.  Due
to the orbital inclination, the distribution function will be mapped
to another function.  Events may then be observed in resonant, wide,
overlap, and isolated configurations.  Although these different
configurations will be detected by different methods, the fact that
they are all drawn from the same distribution means that they all must
be mutually consistent.  Thus, when a sufficient number of one type of
events is measured, one can derive the general distribution function
by calculating the detection efficiency, and then convolving with the
orbital inclination distribution.  Then one can use this 'recovered'
distribution function to predict the number of repeating (overlap,
isolated) events that should have been detected and compare this to
the number that were actually detected (convolved with detection
efficiency) to check that they are mutually consistent.  Subtracting
the number of expected isolated events from the observed distribution
of timescales recovers the true distribution of MACHOs (objects not
associated with stars).

\section{Summary}

\subsection{Repeating Events}

Repeating \ev s  
are a necessary part of the menagerie of \pl-lens \ev s.
These \ev s form a sort of transition or bridge between resonant
and isolated \pl -lens \ev s. 

Individual \rpe s tell us (1) that the lens was very likely to be a \ps;
(2) the projected separation (in units of the \re\ of the \cs );
(3) the mass ratio between the \pl\ and the \cs ; (4) if the \cs\
is luminous, we may be able to learn the mass of the \cs .
The ensemble of repeating \ev s, in combination with resonant \ev s and \is\
\ev s of \shdn, can allow us to extract information about the 
population of planetary systems among the lens population.

\subsection{Planets in Wide Orbits} 

In this paper and its companion 
(\rd\ \& Scalzo 1998) we have introduced the tools needed to systematically use \ml\ to
search for \pl s in wide orbits around distant stars. Because
these searches can be conducted as part of the ongoing \ml\
observations, we have carried out detailed simulations to determine
what signatures should be expected if our Solar System or other
known or model systems serve as lenses. We have also studied how
different detection strategies influence detection efficiencies.
Our results are encouraging in that they clearly indicate that
a systematic search for \pl s in \wo s is not only feasible,
but that, even over the short term,
it can yield interesting results about the population
of \ps s in our own and other galaxies. 

Until now, \ml\ searches for \pl s have concentrated on searching
for \pl s that might be located in the zone for resonant lensing.
Resonant lensing \ev s are expected and will play an important role
in our quest to learn more about \pl s through \ml .
Planetary systems, however, seem likely to exhibit enough structure,
in the form of multiple \pl s, moons revolving about \pl s, and even
belts of compact debris, that it is important to explore all the ways
\ml\ can help us to study them. (See also \rd\ \& Scalzo [1997] and
\rd\ \& Keeton [1998].)   
We point out that \pl s in wide orbits ($a > 1.5 R_E$)
(1) should exist in larger numbers than planets in the \zres ,
(2) also can yield distinctive signatures, and 
(3) may yield larger numbers of detectable \ev s, particularly
if \fsse\ are important. 
Thus, previous estimates of the \ml\ detection rate (see, e.g., Peale 
1997) need to be
revised upward.

The key elements of detectability are to pick up evidence  
of \ml\ \ev s as early as possible, and to be sensitive to \ev s
 even if the peak 
\mage\ is smaller than $1.34.$ 
The detection rate of isolated \pl-lens \ev s is proportional to $n_i$,
where $n_i=w_i/R_{E,i}$, and $w_i$ represents the distance
of closest approach needed for reliable detection of lens $i$.
The rate for repeating events scales as $\prod_{i=1}^k n^{i+1},$ where $k$ is the 
number of repetitions. Just as is the case for the detection of \pl s 
in the \zres , frequent monitoring of ongoing \ev s is important,
although hourly \mo\ may not be necessary.

The features that will improve the rate of \pl\ detection
should be considered when designing the next generation of
\ml\ observations. 
In the meantime, 
the MACHO results to date indicate that the teams have the 
necessary capabilities to detect wide-orbit \pl s (\rd\ \& Scalzo 1998).
 We have suggested a set of modifications
 in the present detection strategy that could, in the short-term,
 significantly
improve detection rates for isolated and repeating \pl-lens \ev s.
One important component is to be alert for the possibility that
repeating \ev s may be \ml\ \ev s, and to continue to
carry out frequent follow-up \mo , even after an
apparently isolated \ev\ has ceased.
It is true that the likelihood of finding a repetition may be 
on the order of a few percent, but the  relative importance
of the \ev s makes the study worthwhile. After all, \ml\ itself
is a low-probability phenomenon, but has been well worth looking for.
In fact, repeating \ev s must be present for a number of 
reasons in addition to the wide-planet connection (Griest \& Hu 1992,
\rd\ \& Mao, 1996).
Nevertheless, 
until recently, the search for them
has remained something of a taboo.    

An important consequence of the likely presence of \pl s in a stellar population
being studied for signs of \ml, is that isolated short \ev s are 
very likely to be present at a level that can be as high (compared to
single stellar-lens \ev s) as ${\cal O}(10\%)$. 
This means that any signature due to low-mass MACHOs cannot be 
unambiguously identified unless the contribution due to \pl s can first be
quantified. (There are exceptions when, for example,
it is known that the majority
of lensing \ev s cannot be due to ordinary stellar systems,
or when the rate of short-time-scale \ev s is so high that the associated
optical depth is larger than could possibly be due to \ps s.) 
Fortunately, it is possible to detect \pl-lens \ev s
which exhibit clear signatures of the fact that the lensing is due to  a
\ps . These distinctive \ev s are repeating and resonant \ev s, and any
short-duration \ev s subject to 
the blending of light from the \cs\ with that of light from the lensed star.
Measuring the rates of such \ev s  
should allow the contribution of \pl s to \shdn, 
apparently isolated \ev s to be quantified.
This contribution can then be subtracted from the total
to derive the magnitude of any contribution due to dark matter existing
outside the realm of 
ordinary stellar systems.   

Another point we have emphasized 
(see especially \rd\ 1998a,b and \rd\ \& Scalzo 1998)
is that, even though the \ps s
\ml\ discovers are far away, some may nevertheless be
the subjects of fruitful further study. Indeed, when the central
star is luminous, we can hope to determine its spectral type.
In some cases, this can help to set the mass scale for the system, and
can therefore
help us to determine the mass of the \pl s that served as lenses.
Finite-source-size effects can also put constraints on the
lens mass.
Thus, although we will not 
image beach front property on the 
\pl s discovered via \ml , 
we should not give up on the possibility of learning more
about individual \ps s that serve as microlenses.
We have also pointed out (\rd\ 1998a) that it is precisely in those systems
in which a \pl\ discovered via \ml\ is most
likely to have Earth-like conditions, that the central star may be
luminous enough to permit further study.

Searches for planets using \ml\ should be able to extend the reach of
local planetary searches by discovering planets in distant parts
of our own and other galaxies and by discovering even low-mass \pl s
orbiting at low speeds. The search for planets in
wide orbits
represents a significant extension of the ongoing microlensing searches.
Indeed, it seems likely that \pl s in \wo s will
provide an important, and
possibly even the dominant mode for the detection of planetary systems
via microlensing, particularly Earth-mass \pl s.

\bigskip
\bigskip
\bigskip
\centerline{ACKNOWLEDGEMENTS}

We would like to thank the referee, Scott Gaudi, for suggested changes to
our original manuscript (\rd\ \& Scalzo 1997); his suggestions
have helped to
improve the presentation.
One of us (RD) would like to thank Andrew Becker,
Arlin Crotts, Andrew Gould, Jean Kaplan, Christopher Kochanek,
David W. Latham, Avi Loeb,
Shude Mao, Robert W. Noyes, Bodhan Paczy\'nski,
Bill Press,
Penny Sackett, Kailash Sahu,
Michael M. Shara, Edwin L. Turner,  Michael S. Turner,
and the participants in the 1997 Aspen workshops, ``The Formation
and Evolution of Planets" and ``Microlensing"
for interesting
discussions, and the Aspen Center for Physics and the Institute for
Theoretical Physics at Santa Barbarba for their hospitality while
this paper was being written,
and the Inter-University Center for Astronomy and Astrophysics in Pune,
India for its hospitality while the paper was revised.
 One of us (RAS) would like to thank the 1996 CfA Summer
Intern Program for support and the Harvard-Smithsonian Center for
Astrophysics for its hospitality while the work was underway.
This work was supported in part by NSF under GER-9450087 and
AST-9619516, and by funding from AXAF.

{}
\end{document}